\documentclass[sn-mathphys-num]{sn-jnl}

\usepackage{lisa}
\usepackage[plain]{fancyref}
\usepackage{tikz}

\usepackage{graphicx}%
\usepackage{multirow}%
\usepackage{amsmath,amssymb,amsfonts}%
\usepackage{amsthm}%
\usepackage{mathrsfs}%
\usepackage[title]{appendix}%
\usepackage{xcolor}%
\usepackage{textcomp}%
\usepackage{manyfoot}%

\theoremstyle{plain}
\newtheorem{theorem}{Theorem}[section]
\theoremstyle{definition}
\newtheorem{definition}{Definition}[section]

\newcommand{\setR}{{\mathord{\mathbb R}}}

\newcommand{\Diff}{\ensuremath{\mathrm{Diff}}}
\newcommand{\GL}{\ensuremath{\mathrm{GL}(n,\mathbb R)}}
\newcommand{\svf}{\ensuremath{\mathfrak{X}}}
\def\ie/{\textit{i}.\textit{e}.}
\def\eg/{\textit{e}.\textit{g}.}
\def\cf/{\textit{cf}.}
\newcommand\restr[2]{{
		\left.\kern-\nulldelimiterspace 
		#1 
		\vphantom{\big|} 
		\right|_{#2} 
}}

\newcommand*{\fancyrefthmlabelprefix}{thm}
\newcommand*{\fancyrefdeflabelprefix}{def}
\newcommand*{\fancyrefeqnlabelprefix}{eqn}
\renewcommand*{\fancyrefdefaultspacing}{\fancyrefloosespacing}
\newcommand*{\Frefdefname}{Definition}
\newcommand*{\frefdefname}{definition}
\frefformat{plain}{\fancyrefdeflabelprefix}{\frefdefname\fancyrefdefaultspacing#1}
\Frefformat{plain}{\fancyrefdeflabelprefix}{\Frefdefname\fancyrefdefaultspacing#1}
\newcommand*{\Frefeqnname}{Equation}
\newcommand*{\frefeqnname}{equation}
\frefformat{plain}{\fancyrefeqnlabelprefix}{\frefeqnname\fancyrefdefaultspacing#1}
\Frefformat{plain}{\fancyrefeqnlabelprefix}{\Frefeqnname\fancyrefdefaultspacing#1}
\newcommand*{\Frefthmname}{Theorem}
\newcommand*{\frefthmname}{theorem}
\frefformat{plain}{\fancyrefthmlabelprefix}{\frefthmname\fancyrefdefaultspacing#1}
\Frefformat{plain}{\fancyrefthmlabelprefix}{\Frefthmname\fancyrefdefaultspacing#1}

\renewcommand{\fref}{\Fref}

\usepackage[disable]{todonotes}

\usepackage{hyperref}
\hypersetup{
    hypertexnames=false,
    colorlinks,
    linkcolor={red!50!black},
    citecolor={blue!50!black},
    urlcolor={blue!80!black}
}

\usetikzlibrary{cd}

\begin{document}

\title
{A Construction of the Lie Algebra of a Lie Group in Isabelle/HOL}


\author*[1]{\fnm{Richard} \sur{Schmoetten}}\email{richard.schmoetten@ed.ac.uk}

\author[1]{\fnm{Jacques D.} \sur{Fleuriot}}\email{jdf@ed.ac.uk}


\affil[1]{\orgdiv{School of Informatics}, \orgname{University of Edinburgh}, \orgaddress{\street{10 Crichton Street}, \city{Edinburgh}, \postcode{EH8 9AB}, \country{Scotland, UK}}}




\abstract{%
This paper describes a formal theory of smooth vector fields, Lie groups and the Lie algebra of a Lie group in the theorem prover Isabelle.
Lie groups are abstract structures that are composable, invertible and differentiable. They are pervasive as models of continuous transformations and symmetries in areas from theoretical particle physics, where they underpin gauge theories such as the Standard Model, to the study of differential equations and robotics. 
Formalisation of mathematics in an interactive theorem prover, such as Isabelle, provides strong correctness guarantees by expressing definitions and theorems in a logic that can be checked by a computer.
Many libraries of formalised mathematics lack significant development of textbook material beyond undergraduate level,
and this contribution to mathematics in Isabelle aims to reduce that gap, particularly in differential geometry. 
We comment on representational choices and challenges faced when integrating complex formalisations, such as smoothness of vector fields, with the restrictions of the simple type theory of HOL.
This contribution paves the way for extensions both in advanced mathematics, and in formalisations in natural science.
}




\maketitle

\section{Introduction}\label{sec:intro}
The development of mathematical foundations for physical and other scientific theories is the central content of most theoretical sciences.
Mathematics is both regarded as highly rational, so that mathematical formulation implies thorough understanding of a phenomenon,
and often has algorithmic or computational applications,
so that mathematical theories have many uses \eg/ in predicting or simulating natural phenomena.
A lofty ideal of this first aspect is a further formalisation not just in terms of pen-and-paper mathematics, but a formal, computer-checked logical framework, eliminating much of the potential for undiscovered errors and unconsidered edge cases.
However, given the mathematical refinement of many of the most advanced theories of natural science, this requires considerable formalisation effort even just in pure mathematics: theoretical physics, for example, has been studied for millennia, while formalisation of mathematics is a mere few decades old.
The purpose of this paper is to present a computer-checked formalisation of mathematical material in the domains of linear algebra, differential geometry and Lie groups.
Lie theory is particularly interesting, as it is ubiquitous in science: Lie groups model continuous transformations, from rotations to conformal mappings \cite{schottenloher2008}.
Special Lie groups called the Lorentz and Poincar\'{e} groups specify symmetries in special and general relativity \cite{stephani2004}; 
the representation theory of Lie groups and their algebras underlies modern particle physics \cite{woit2017}.

This paper is a contribution to the field of formal mathematics in its own right, discussing the representation of advanced mathematical concepts in the simple type theory of Isabelle/HOL, but it also aims to provide some of the necessary context for formalisation in the natural sciences.

Set- and locale-based formalisations in the simple type theory of Isabelle/HOL have gained traction in recent efforts to formalise complex mathematical definitions that deal with sets of structures (\eg/ direct limits), substructures (\eg/ subgroups), and structure-preserving maps (\eg/ homomorphisms).
Examples include sheaves of rings (and topology) \cite{bordg2022}, smooth manifolds (and linear algebra) \cite{immler2019}, and Stone Algebras \cite{guttmann2020}.
The interaction between classical type-based definitions (such as functions, charts and Euclidean spaces) and set-based ones (such as manifolds and tangent spaces) is consequently one of the main topics of discussion arising from our formalisation work.


We begin with a short account of pre-existing theories in \fref{sec:background}, introducing a mixture of textbook and mechanised definitions in differential geometry to give the reader some familiarity both with Isabelle, and the mathematical prerequisites.
We then present our own formalisation of topics in the same field, including the diffeomorphism group of a manifold (\fref{sec:diff}) and smooth vector fields (\fref{sec:svf}).
These lead up to a formal theory of Lie groups in \fref{sec:lie}, including some concrete Lie groups, and both an abstract definition of \emph{a} Lie algebra and a construction of \emph{the} Lie algebra of a Lie group, which relies on most of the preceding work.

\section{Background}\label{sec:background}
Isabelle is an interactive theorem prover in the LCF tradition \cite{paulson1989,paulson1990,isar-ref}: a small kernel of trusted propositions and rules can be combined to construct new theorems, which in turn become useable in further theorem constructions.
Isabelle provides a logical framework (or meta-logic) in which logics of interest can be formalised.
Higher-order logic (HOL) has emerged as the most convenient and best-developed such object logic; we therefore use Isabelle/HOL for the entire contribution described in this paper.
In addition to the tools distributed as standard with Isabelle, we use a contribution by Unruh\footnote{accessible from Unruh's github repository \href{https://github.com/dominique-unruh/isabelle-tidbits/tree/1d00f7f1f2fe76df49f6caeb35d83d2e091470bf/Wlog}{(link)}} to enable reasoning ``without loss of generality'', for clearer proof structures.

Our formalisation, like that of Immler and Zhan (which we review in \fref{sec:manifolds-review}) \cite{immler2019}, is largely locale-based, and we will often compare locales and types (or type classes).
A \emph{locale} in Isabelle/HOL can be seen as a parametrised context, which allows for local assumptions while leaving the global logic unaffected \cite{kammuller1999,ballarin2014}. We recommend Ballarin's exploration of algebraic locales as a topical introduction \cite{ballarin2020} beyond our own \fref{sec:locales}. 
A \emph{type class} is a collection of types, such that each type in the class satisfies a given proposition (sometimes called the \emph{sort constraint}), and provides an implementation of a given constant (\eg/ an element or operation, which may occur in the sort constraint) \cite{haftmann2007,holzl2013}.
Type classes are used to provide polymorphic constants with axiomatised behaviour through type annotations, but simple type theory limits the complexity of these constraints.
In Isabelle, type variables are written with a leading apostrophe, \eg/ \lisa|'var|.
Double colons are used both to annotate a type variable with class requirements, and object variables with type information, \eg/ \lisa[keywordstyle={[2]}]|'var::class|.

\subsection{Locales and classes for algebra in HOL}\label{sec:locales}

This section presents some common mathematical definitions relevant to differential geometry. We use formalisations of these concepts, drawn from our own work as well as Immler and Zhan's \cite{immler2019}, to showcase the flexibility and expressiveness of Isabelle's locales, and their use alongside algebraic theories developed for type classes.

Consider, as a first example, a mechanisation of modules (over rings).
Isabelle comes with a set of \emph{syntactic type classes}, whose only purpose is to provide notation for functions with a given signature.
One such class is \lisa|plus|; any type \lisa|'a::plus| in this class must provide a function denoted \lisa|+| with signature \lisa|'a \<Rightarrow> 'a \<Rightarrow> 'a|.
The type class \lisa|comm_ring| is a subclass of the syntactic classes \lisa|plus| and \lisa|mult| (which provides \lisa|*|), and imposes axioms on these operations to give them the usual meaning of commutative ring addition and multiplication, respectively.
The class definition below should be seen as an example only: the standard definition of this class in Isabelle is integrated into a much richer hierarchy of classes, and looks different as a result. Omissions in example listings are marked by an ellipsis, omitted proofs of listed statements are marked as \lisa|<proof>|. 

\begin{lstlisting}
class comm_ring = plus + mult + (*...*)
  assumes distrib: "(a + b) * c = a * c + b * c"
    and (*...*)
\end{lstlisting}

We can now work inside abstract commutative rings by annotating type variables with the class \lisa|comm_ring|.
For example, a set $S$ with elements in a commutative ring can be written \lisa|S :: ('a::comm_ring) set|.
\todo{mention type inference?}The convenience of writing theorems about rings with so little verbosity, particularly when type annotations can often be omitted in the presence of Isabelle's type inference, is balanced by the limitations imposed on axioms (or \emph{sort constraints}) of type classes by the simple type system of Isabelle/HOL: in particular, only a single type variable may appear in sort constraints.
To circumvent this issue (and others), we use locales.

A module is a set $S$ such that any two elements of $S$ can be added together to produce some other element of $S$, and can be scaled by an arbitrary element of some fixed ring $R$, again to produce an element of $S$.
Notice this involves two type variables: \lisa|'b| (with addition) for elements of $S$, and \lisa|'a| (a commutative ring with identity) for $R$. 

\begin{lstlisting}
locale module_on =
  fixes S
    and scale :: "'a::comm_ring_1 \<Rightarrow> 'b::ab_group_add \<Rightarrow> 'b" (infixr "*s" 75)
  assumes mem_add: "\<lbrakk>x \<in> S; y \<in> S\<rbrakk> \<Longrightarrow> x + y \<in> S"
    and mem_scale: "x \<in> S \<Longrightarrow> a *s x \<in> S"
    and (*...*)
\end{lstlisting}

This locale definition, adapted from Immler and Zhan \cite[Sec.~4.4]{immler2019}, can be read as ``\lisa|scale| (denoted \lisa|*s|) defines a module on \lisa|S|'', and presents several differences to our earlier type class definition.
The first is the presence of two type variables: as mentioned above, this is not possible for classes.
This issue is side-stepped by the main library in Isabelle/HOL by defining classes only for modules (and vector spaces) over the real numbers; \ie/ the type variable \lisa|'a| above is replaced with a concrete type \lisa|real|.
Secondly, instead of using a syntactic class, we pass the scaling function directly to the locale definition as a parameter. The properties of this function are axiomatised only on a set \lisa|S|, given also as a locale parameter.
This allows for easier formulation of theorems about submodules: one can simply consider the \lisa|module_on| formed by a suitable subset of \lisa|S| under the same operation \lisa|*s|.
In contrast, talking about submodules using the type class is tricky, because one would need to carve out a new type for any submodule, and HOL has no explicit way of talking about ``subtypes''. \todo{improve this last sentence/argument}

A more complicated example of a locale definition
is the module homomorphism: a structure-preserving map between two modules over the same ring. Similar morphisms will be mentioned throughout this paper for many other locale-based structures. All of them instantiate two locales of the same kind (modules, in this case), and often name these instances (\lisa|m1| and \lisa|m2| here). They then impose axioms to relate operations of these two similar structures (\eg/ scaling \lisa|*a| and \lisa|*b| in the case of modules) through a fixed function \lisa|f|. Examples include diffeomorphisms in \fref{sec:manifolds-review}; linear isomorphisms (module homomorphisms between vector spaces) in \fref{sec:tangent-coord}; and group homomorphisms (which satisfy the axiom \lisa|add| below) in \fref{sec:action}.
\begin{lstlisting}
locale module_hom_on = m1: module_on S1 s1 + m2: module_on S2 s2
  for S1 :: "'b::ab_group_add set" and S2 :: "'c::ab_group_add set"
    and s1 :: "'a::comm_ring_1 \<Rightarrow> 'b \<Rightarrow> 'b" (infixr "*a" 75)
    and s2 :: "'a::comm_ring_1 \<Rightarrow> 'c \<Rightarrow> 'c" (infixr "*b" 75) +
  fixes f :: "'b \<Rightarrow> 'c"
  assumes add: "\<And>x y. \<lbrakk>x \<in> S1; y \<in> S1\<rbrakk> \<Longrightarrow> f (x + y) = f x + f y"
    and scale: "\<And>x. x \<in> S1 \<Longrightarrow> f (r *a x) = r *b (f x)"
\end{lstlisting}

The instantiation \lisa|m1: module_on S1 s1| above, stating that \lisa|S1| is a module under \lisa|s1|, has a counterpart outside locale definitions.
Abstract locales can be interpreted by structures that satisfy their axioms.
This is done with the Isabelle command \lisa|interpretation|, which requires a proof of the relevant locale axioms for the structures provided.
For example, the set of real numbers, written as \lisa|{x::real. True}| in Isabelle, is a module under multiplication \lisa[commentstyle=]|(*)|, and interprets the locale \lisa|module_on|.
\begin{lstlisting}[commentstyle=]
interpretation R: module_on "{x::real. True}" "(*)"
  by (simp add: module_on_def ring_class.ring_distribs)
\end{lstlisting}
This interpretation gives access to all lemmas shown for \lisa|module_on|, specific to real numbers, \eg/ a lemma named \lisa|foo| inside the locale \lisa|module_on| will become available under the name \lisa|R.foo|, with the generic parameters \lisa|S, scale| of the module replaced by the set of real numbers and multiplication.
The \lisa|sublocale| command can be used similarly to instantiate a locale from within another locale.

We give some further examples of standard definitions from linear algebra here, which will be needed later.
A vector space is easily defined as a module where the underlying ring is in fact a \lisa|field| (\ie/ it has multiplicative inverses). The locale definition looks just as simple: we merely adjust parameters to have the desired type, and extend the locale \lisa|module_on|.
\begin{lstlisting}
locale vector_space_on = module_on S scale
  for S and scale :: "'a::field \<Rightarrow> 'b::ab_group_add \<Rightarrow> 'b" (infixr "*s" 75)
\end{lstlisting}
A similar procedure yields linear maps between vector spaces; the type of the function \lisa|f| can be inferred from the definition of \lisa|module_hom_on|.
\begin{lstlisting}
locale linear_on = module_hom_on S1 S2 s1 s2 f
  for S1 and s1::"'a::field \<Rightarrow> 'b \<Rightarrow> 'b::ab_group_add"
    and S2 and s2::"'a::field \<Rightarrow> 'c \<Rightarrow> 'c::ab_group_add"
    and f
\end{lstlisting}

\todo{We could use locales to define everything, but it's good to use existing theory where possible}



A map $f: V \times W \to X$ between a pair of vector spaces $V \times W$ and a third vector space $X$ is bilinear if it is linear as a map $V \to X$ (the locale name doubles as a predicate: \lisa|linear_on V X|) and $W \to X$ for arbitrary fixed elements of $W$ and $V$ respectively.
We omit some type annotations, but note that $V, W, X$ are subsets of (potentially different) types \lisa|'b|, \lisa|'c|, \lisa|'d|
in the type class of abelian groups. 
The set $X$, for example, additive by virtue of its type class, is made into a \lisa|vector_space_on| by the scaling operation \lisa|scaleX|, where the field is implicit in the type \lisa|'a::field| of \lisa|scaleX|.

%

\begin{lstlisting}
locale bilinear_on =
    vector_space_pair_on V W scaleV scaleW +
    vector_space_on X scaleX
  for V (*...*) and W (*...*) and X::"'d::ab_group_add set"
    and scaleV (*...*) and scaleW (*...*) and scaleX::"'a::field\<Rightarrow>'d\<Rightarrow>'d" +
  fixes f::"'b\<Rightarrow>'c\<Rightarrow>'d"
  assumes linearL: "w\<in>W \<Longrightarrow> linear_on V X scaleV scaleX (\<lambda>v. f v w)"
    and linearR: "v\<in>V \<Longrightarrow> linear_on W X scaleW scaleX (\<lambda>w. f v w)"
\end{lstlisting}
An algebra is then a vector space on $S$ with a bilinear multiplication $S \times S \to S$. Notice in the above that \lisa|vector_space_on| is also a locale: showing a structure is a model of \lisa|bilinear_on| will require a proof that the relevant set \lisa|X| is a vector space. In exchange, results about the vector space \lisa|X| are available as facts inside the locale for bilinear maps.

\subsection{Smooth manifolds in Isabelle/HOL -- a brief review}\label{sec:manifolds-review}

We now give a brief review of Immler and Zhan's central ideas \cite{immler2019}, and examine how design choices relating to partiality impact the work described in this paper; Isabelle listings in that section are reproduced from their AFP entry\footnote{The Archive of Formal Proofs: \url{https://www.isa-afp.org/entries/Smooth_Manifolds.html}}.
For more detail on the formalisation, we refer to the paper above; good textbooks expositions of differential geometry include works by Lee \cite{lee2012} (which informs Immler and Zhan's formalisation as well as ours), and Spivak \cite{spivak1999}.

An $n$-dimensional manifold is a topological space that is locally Euclidean \cite[Chap.~1]{lee2012}.
Formally, it is defined both by its topology, which we require to be second-countable and Hausdorff \cite[App.~A]{lee2012}, and by a set of \emph{charts}, which are local homeomorphisms from some domain in $M$ to a codomain in $\setR^n$.

Isabelle has several theories of topology; Immler and Zhan use the type class \lisa|topological_space|.
Any type \lisa|'a| in this class comes equipped with a predicate \lisa|open :: 'a set \<Rightarrow> bool|, such that open sets satisfy the usual topological axioms.
\begin{lstlisting}
class topological_space = "open" +
  assumes open_UNIV: "open UNIV"
  assumes open_Int: "\<lbrakk>open S; open T\<rbrakk> \<Longrightarrow> open (S \<inter> T)"
  assumes open_Union: "\<forall>S\<in>K. open S \<Longrightarrow> open (\<Union>K)"
\end{lstlisting}
This follows the same principle we saw in \fref{sec:locales}, where the syntax \lisa|+| is subjected to an invariant by the class \lisa|comm_ring|.
Topological spaces are said to be Hausdorff (or $\mathrm{T}2$) if they admit disjoint open neighbourhoods around distinct points, and second-countable if they have a countable (topological) basis.
These conditions are a bit technical, and we omit listings of the subclasses \lisa|t2_space| and \lisa|second_countable_topology|.
In general, they make our topological spaces a little more similar to our well-known Euclidean space: for example, convergent sequences in T2 spaces have unique limits.
Immler and Zhan use second-countability to construct partitions of unity, and those partitions allow us to obtain suitable extensions of functions when going between tangent spaces (defined later in this section).
This will be discussed in \fref{sec:tangent-coord}, see also \fref{fig:tangent-space-maps}.

A homeomorphism is a continuous bijection with continuous inverse.
Notice that while the topology is a global property of the type, \lisa|homeomorphism| is defined explicitly on a domain set \lisa|S| and codomain set \lisa|T|. The functions \lisa|f, g| that implement this homeomorphism may not be continuous or mutually inverse outwith \lisa|S, T|.
\begin{lstlisting}
definition "homeomorphism S T f g \<longleftrightarrow>
  (\<forall>x\<in>S. (g(f x) = x)) \<and> (f ` S = T) \<and> continuous_on S f \<and>
  (\<forall>y\<in>T. (f(g y) = y)) \<and> (g ` T = S) \<and> continuous_on T g"
\end{lstlisting}
We also give Immler and Zhan's definition of a chart as a type constructor that takes a topological type \lisa|'a| and a Euclidean type \lisa|'e| as inputs, and creates a new type whose elements are homeomorphisms (quadruples satisfying the predicate \lisa|homeomorphism| above) between open sets.
\begin{lstlisting}
typedef (overloaded) ('a::topological_space, 'e::euclidean_space) chart =
  "{(d::'a set, d'::'e set, f, f').
    open d \<and> open d' \<and> homeomorphism d d' f f'}"
\end{lstlisting}

Differential geometry aims to extend calculus from Euclidean spaces to the more general manifolds.
We now clarify some vocabulary for the usual Euclidean calculus.
A function $f$ between two Euclidean spaces will be called $k$-\emph{smooth} at a point $x$ if it has a continuous $k$-th derivative at $x$.
When $f$ is infinitely differentiable, we omit $k$ and simply say $f$ is smooth (both in prose and in Isabelle); $0$-smoothness is equivalent to continuity.
Smoothness and higher differentiability are formalised over an explicit set in Isabelle, so that \lisa|k-smooth_on S f| means $f$ is $k$-smooth at each point in the set $S$. Notice that $k$ is an extended natural number (it may be infinite)\footnote{%
    The theory \lisa|Extended_Nat| distributed with Isabelle introduces a type that includes both the natural numbers and a special element \lisa|\<infinity>|. This is reminiscent to the \lisa|option| type constructor we employ elsewhere: exactly one distinguished value is added to the existing type \lisa|nat|. See \url{https://isabelle.in.tum.de/library/HOL/HOL-Library/Extended_Nat.html}.
},
and higher differentiability (\lisa|higher_differentiable_on|) of $f$ is defined recursively over a natural number $n$, where the base case $n=0$ is continuity, and for $n\geq1$ the derivative of $f$ exists and is $(n-1)$-higher differentiable. 
\begin{lstlisting}
definition k_smooth_on ("_-smooth'_on" [1000]) where
 smooth_on_def: "k-smooth_on S f = (\<forall>n\<le>k. higher_differentiable_on S f n)"
\end{lstlisting}

An $n$-dimensional manifold $M$ (without boundary) is defined by a \emph{set} of such charts.
The reasoning follows much the same pattern as in \fref{sec:locales}: we will need to talk about submanifolds a lot when dealing with individual charts as well as tangent spaces (see \fref{sec:manifolds-review}), so sets will be simpler to handle than types.
$M$ is a $C^k$-manifold, or $k$-smooth, if, whenever two charts $\phi, \psi$ overlap, their compositions $\phi \circ \psi^{-1}$ and  $\psi \circ \phi^{-1}$ are $k$-smooth as maps on the Euclidean space.

The predicate \lisa|k-smooth_compat| captures this property of two charts \lisa|c1| and \lisa|c2| (type annotation omitted).
\begin{lstlisting}
definition smooth_compat ("_-smooth'_compat" [1000])
  where "smooth_compat k c1 c2 \<longleftrightarrow>
    (k-smooth_on (c1 ` (domain c1 \<inter> domain c2)) (c2 \<circ> inv_chart c1) \<and>
     k-smooth_on (c2 ` (domain c1 \<inter> domain c2)) (c1 \<circ> inv_chart c2) )"
\end{lstlisting}

For the rest of this paper, we fix $k = \infty$: this is a common axiom for Lie groups, and allows the construction of useful isomorphisms between tangent spaces (defined later in this section), which we rely on in \fref{sec:TM}.
Variants of the material in this section for arbitrary $k$, whenever useful, can be found both directly in the formalised theories, and in mathematical literature \cite{lee2012}.

\begin{lstlisting}
locale smooth_manifold =
  fixes charts::"('a::{second_countable_topology, t2_space},
                  'e::euclidean_space) chart set"
  assumes pairwise_compat:
    "\<lbrakk>c1 \<in> charts; c2 \<in> charts\<rbrakk> \<Longrightarrow> \<infinity>-smooth_compat c1 c2"
\end{lstlisting}
The maximal set of all charts obeying this condition is the \emph{atlas} of $M$.
The set of points contained in a manifold is the union of the charted domains%
\footnote{One could alternatively require that the entire topological space is covered by charts. In Isabelle, this would make submanifolds more awkward, since the underlying topology is always defined on the entire type.}:
\[M = \bigcup_{\phi \in \mathrm{atlas}} \mathrm{Dom}(\phi).\]
Mathematical textbooks mildly abuse notation and conflate the manifold $M$ with its set of points, and in the current presentation we shall do so when opportune.
A trivial, yet important, example of a manifold is any Euclidean space itself, with the identity as the chart. More models can be found in \fref{sec:lie-models}.

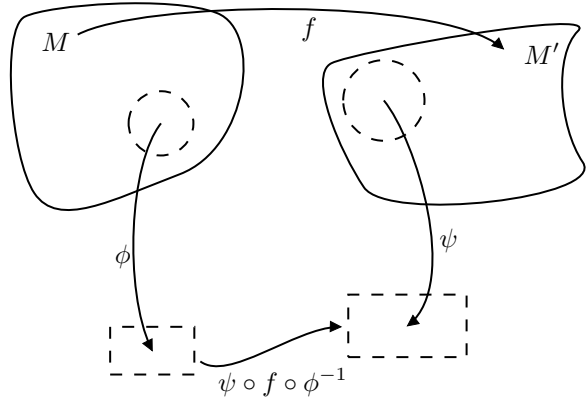
\begin{figure}
\centering
\tikzset{every picture/.style={line width=0.75pt}} 

\begin{tikzpicture}[x=0.75pt,y=0.75pt,yscale=-.6,xscale=.6]

\draw   (123.6,49.6) .. controls (143.6,39.6) and (245.6,31.6) .. (283.6,46.6) .. controls (321.6,61.6) and (309.6,159.6) .. (252.6,181.6) .. controls (195.6,203.6) and (151.6,227.6) .. (131.6,197.6) .. controls (111.6,167.6) and (103.6,59.6) .. (123.6,49.6) -- cycle ;
\draw   (378.6,88.6) .. controls (398.6,78.6) and (603.6,41.6) .. (583.6,61.6) .. controls (563.6,81.6) and (567.6,141.6) .. (587.6,171.6) .. controls (607.6,201.6) and (425.6,222.6) .. (405.6,192.6) .. controls (385.6,162.6) and (358.6,98.6) .. (378.6,88.6) -- cycle ;
\draw  [dash pattern={on 4.5pt off 4.5pt}] (209.96,138.3) .. controls (210.34,123.42) and (222.72,111.67) .. (237.6,112.06) .. controls (252.47,112.45) and (264.22,124.82) .. (263.83,139.7) .. controls (263.44,154.57) and (251.07,166.32) .. (236.19,165.93) .. controls (221.32,165.55) and (209.57,153.17) .. (209.96,138.3) -- cycle ;
\draw  [dash pattern={on 4.5pt off 4.5pt}] (388.4,119.8) .. controls (388.4,101.13) and (403.53,86) .. (422.2,86) .. controls (440.87,86) and (456,101.13) .. (456,119.8) .. controls (456,138.47) and (440.87,153.6) .. (422.2,153.6) .. controls (403.53,153.6) and (388.4,138.47) .. (388.4,119.8) -- cycle ;
\draw    (236.89,139) .. controls (210.3,172.49) and (205.73,278.36) .. (228.54,327.4) ;
\draw [shift={(229.6,329.6)}, rotate = 243.43] [fill={rgb, 255:red, 0; green, 0; blue, 0 }  ][line width=0.08]  [draw opacity=0] (8.93,-4.29) -- (0,0) -- (8.93,4.29) -- cycle    ;
\draw  [dash pattern={on 4.5pt off 4.5pt}] (194.6,309.6) -- (264.6,309.6) -- (264.6,349.6) -- (194.6,349.6) -- cycle ;
\draw  [dash pattern={on 4.5pt off 4.5pt}] (392.6,282.6) -- (491,282.6) -- (491,335) -- (392.6,335) -- cycle ;
\draw    (422.2,119.8) .. controls (449.19,153.09) and (485.3,271.77) .. (443.75,307.24) ;
\draw [shift={(441.8,308.8)}, rotate = 323.17] [fill={rgb, 255:red, 0; green, 0; blue, 0 }  ][line width=0.08]  [draw opacity=0] (8.93,-4.29) -- (0,0) -- (8.93,4.29) -- cycle    ;
\draw    (269.6,338.6) .. controls (293,356.15) and (347.77,310.96) .. (383.86,309.62) ;
\draw [shift={(386.6,309.6)}, rotate = 181.59] [fill={rgb, 255:red, 0; green, 0; blue, 0 }  ][line width=0.08]  [draw opacity=0] (8.93,-4.29) -- (0,0) -- (8.93,4.29) -- cycle    ;
\draw    (167.6,64.6) .. controls (219.08,36.88) and (466.58,31.7) .. (520.04,75.26) ;
\draw [shift={(521.6,76.6)}, rotate = 221.99] [fill={rgb, 255:red, 0; green, 0; blue, 0 }  ][line width=0.08]  [draw opacity=0] (8.93,-4.29) -- (0,0) -- (8.93,4.29) -- cycle    ;

\draw (466,225) node [anchor=north west][inner sep=0.75pt]   [align=left] {$\displaystyle \psi $};
\draw (196,238) node [anchor=north west][inner sep=0.75pt]   [align=left] {$\displaystyle \phi $};
\draw (135,61) node [anchor=north west][inner sep=0.75pt]   [align=left] {$\displaystyle M$};
\draw (536,69) node [anchor=north west][inner sep=0.75pt]   [align=left] {$\displaystyle M'$};
\draw (282,342) node [anchor=north west][inner sep=0.75pt]   [align=left] {$\displaystyle \psi \circ f\circ \phi ^{-1}$};
\draw (351,48) node [anchor=north west][inner sep=0.75pt]   [align=left] {$\displaystyle f$};

\end{tikzpicture}
\caption{\label{fig:differentiability}%
  Differentiability of $f\colon M \to M'$ is defined using the Euclidean function $\psi \circ f \circ \phi^{-1}$.
  Chart (co)domains are drawn in dashed lines. See \fref{def:diff}.
}
\end{figure}

\begin{definition}[Differentiability on Manifolds; Diffeomorphism]\label{def:diff}
A function $f\colon M \to M'$ between two smooth manifolds is smooth at a point $p \in M$ if there are charts $\phi, \psi$ (in the atlas of $M$ and $M'$ respectively) such that
\begin{enumerate}
\item $p\in\mathrm{Dom}(\phi)$,
\item $f(\mathrm{Dom}(\phi)) \subseteq \mathrm{Dom}(\psi)$, and
\item $\psi \circ f \circ \phi^{-1}$ is smooth on $\mathrm{Codom}(\phi)$ (as a function between Euclidean spaces).
\end{enumerate}
If this holds for all points in $M$, then $f$ is smooth on $M$.
A \emph{diffeomorphism} is a smooth bijection with a smooth inverse. 
See \fref{fig:differentiability} for an illustration of smoothness at a point.
\end{definition}

For legibility, we give the locale corresponding to this definition, \lisa|diff|, under our stated condition $k=\infty$, and omit the \lisa|for| clause which contains the usual type annotations for charts.
Naming the two instances of \lisa|smooth_manifold| as \lisa|src| (for source) and \lisa|dest| (for destination) gives us easy access to the respective atlases and carrier sets.
The function \lisa|inv_chart| returns the inverse of a chart (as a function)\footnote{Whenever a chart is unambiguously used as a function, the \lisa|chart| is silently coerced into its defining function \lisa|'a\<Rightarrow>'e|.}; \lisa|domain| and \lisa|codomain| similarly return the domain and codomain of a chart.

\begin{lstlisting}
locale diff = src: smooth_manifold charts1 + dest: smooth_manifold charts2 (*...*) +
 fixes f :: "'a \<Rightarrow> 'b"
 assumes exists_smooth_on: "x \<in> src.carrier \<Longrightarrow>
   \<exists>c1\<in>src.atlas. \<exists>c2\<in>dest.atlas.
     x \<in> domain c1 \<and>
     f ` domain c1 \<subseteq> domain c2 \<and>
     smooth_on (codomain c1) (c2 \<circ> f \<circ> inv_chart c1)"
\end{lstlisting}
The locale \lisa|diffeomorphism| is defined by an inverse pair of functions, both of which must be smooth (\ie/ they must interpret the locale \lisa|diff| for $k=\infty$). 
The listing below makes use of namespaces for sublocales and interpretations: \lisa|inv| collects results about the differentiable function $f'$ (on the carrier set, $f' = f^{-1}$), while \lisa|src| and \lisa|dest| refer to the \lisa|smooth_manifold| defined by the charts \lisa|charts1| and \lisa|charts2| respectively.\todo{vocab: namespacing? interpretation/sublocale?}

\enlargethispage{-1\baselineskip}
\begin{lstlisting}
locale diffeomorphism = diff \<infinity> charts1 charts2 f + inv: diff \<infinity> charts2 charts1 f'
  for charts1 charts2 f f' +
  assumes f_inv: "\<And>x. x \<in> src.carrier \<Longrightarrow> f' (f x) = x"
    and f'_inv:  "\<And>y. y \<in> dest.carrier \<Longrightarrow> f (f' y) = y"
\end{lstlisting}

The vector space of smooth real-valued functions (under pointwise addition and scaling) on $M$ is denoted $C^\infty(M)$.
A linear map $v\colon C^\infty \to \setR$ is a \emph{derivation} at $p$ (\lisa|is_derivation v p|) if it obeys
\begin{equation}\label{eqn:leibniz}
v(f \circ g) = f(p)v(g) + v(f)g(p).\tag{Leibniz rule}
\end{equation}
The set $T_pM$ of all derivations at $p$ is a real $n$-dimensional vector space under the usual pointwise operations. It is called the \emph{tangent space} at $p$, and its elements are called \emph{tangent vectors} (at $p$).
Tangent vectors are local: they act identically on functions in $C^\infty(M)$ whenever those functions are indistinguishable in some neighbourhood of $p$.%
\footnote{%
  In fact, one could define tangent vectors through their action on \emph{germs} of functions at $p$ instead, \ie/ equivalence classes of functions with identical behaviour in a neighbourhood of $p$ \cite[p.~71]{lee2012}.%
}
Any smooth map between manifolds can be used to \emph{push forward} tangent vectors from its domain to its codomain.
\begin{definition}[Push-forward]
Let $F$ be a smooth map between manifolds $M$ and $N$.
Then we define the \emph{push-forward} of $F$ at $p \in M$:
\begin{alignat*}{3}
& dF_p \colon & T_pM \to \: & T_{F(p)}N \\
&& dF_p (v) = \: & f \mapsto v(f \circ F)
\end{alignat*}
This is a linear map, and if $F$ is a diffeomorphism then $dF_p$ is an isomorphism of tangent spaces, with $(dF_p)^{-1} = d(F^{-1})_{F(p)}$.
We often omit the subscript $p$ when the point is either arbitrary or clear from context.\todo{This is a bit disingenuous, as dropping $p$ is meaningful: the differential is a functor from the category of manifolds to the category of smooth vector bundles. But that's abstract nonsense at this point. It does mean that we can forget about $p$ and consider vector fields instead of vectors. Footnote? See problem 10-3 in Lee.}
\end{definition}

Many total functions of type \lisa|'a\<Rightarrow>real| implement the same partial function on the manifold -- behaviour outside the manifold's carrier set is unconstrained.
To obtain a unique representative for each smooth function, Immler and Zhan include the additional condition that elements of $C^\infty(M)$ (\lisa|diff_fun_space|) be zero outside the carrier set of the manifold (\lisa|extensional0|).
This choice is carried forward to the definition of tangent spaces.

\vbox{\begin{lstlisting}
definition "extensional0 A f = (\<forall>x. x \<notin> A \<longrightarrow> f x = 0)"

definition diff_fun_space :: "('a \<Rightarrow> real) set" where
  "diff_fun_space = {f. diff_fun k charts f \<and> extensional0 carrier f}"

definition tangent_space :: "'a \<Rightarrow> (('a \<Rightarrow> real) \<Rightarrow> real) set" where
  "tangent_space p = {X. is_derivation X p \<and> extensional0 diff_fun_space X}"
\end{lstlisting}}
Overall, because of the central importance of vector spaces in the theory of manifolds, the predicate \lisa|extensional0| has many advantages, \eg/ the \lisa|tangent_space| is a vector space under the pointwise operations of addition \lisa|(+)| and real scaling \lisa[commentstyle=]|(*\<^sub>R)| defined on the whole function type (\cf/ \fref{sec:locales}).
However, it also means that the tangent space at a point $p$ (as well as $C^\infty(M)$) depends globally on the manifold on which it is defined: if $U$ is a strict submanifold of $M$, and $p \in U$, then $T_pU \neq T_pM$.
Note we still have a linear isomorphism $T_pU \simeq T_pM$, as shown in \fref{fig:tangent-space-maps}.

\begin{figure}
\centering
\[\begin{tikzcd}
	U &&&& {\psi(U)} && {T_{\psi(p)}\psi(U)} && {T_pU} \\
	M &&&& {\mathbb R^n} && {T_{\psi(p)}\mathbb R^n} && {T_pM}
	\arrow["\iota"', from=1-1, to=2-1]
	\arrow["d\iota", from=1-9, to=2-9]
	\arrow["d\kappa"', from=1-7, to=2-7]
	\arrow["{\psi :: \texttt{('a,'e)chart}}"{description}, from=1-1, to=1-5]
	\arrow["\nabla"{description}, from=2-5, to=2-7]
	\arrow["d\psi"', from=1-9, to=1-7]
	\arrow["\kappa", from=1-5, to=2-5]
\end{tikzcd}\]
\caption{\label{fig:tangent-space-maps}%
  Let the type \lisa|'e::euclidean_space| implement $\setR^n$. Let $\psi$ be a chart of $M$ with domain $U$, and consider inclusions $\iota : U \hookrightarrow M$ and $\kappa : \psi(U) \hookrightarrow \setR^n$.
  Then the push-forward maps $d\iota$, $d\kappa$ and $d\psi$ are linear isomorphisms of tangent spaces.
  The directional derivative $\nabla: y \mapsto (y \cdot \restr \nabla x)$ implements the isomorphism between Euclidean space and its tangent space at $x$.
}
\end{figure}

Since we base our development on the theory of Immler and Zhan, we therefore have to carry explicit conditions forward through our definitions, including the diffeomorphism group (\fref{sec:diff}), smooth vector fields, and their action on function spaces (\fref{sec:derivations}).
In contrast, the topology in which the charts are homeomorphisms is encoded in the type class of the type \lisa|'a|, and therefore must be defined on the entire type universe.
This discrepancy between the \emph{type-class} topology of the manifold, and the \emph{set} of charts defining it, has repercussions on our work with vector fields in \fref{sec:TM}.

A major theorem of Immler and Zhan's work formalises that $\mathrm{Dim}(T_pM) = \mathrm{Dim}(M) = \mathrm{Dim}(\setR^n)$.
They show the maps $\nabla$, $d\iota$, $d\kappa$ and $d\psi$ of \fref{fig:tangent-space-maps} are linear bijections, and use the fact such maps preserve the dimension of a vector space.
They obtain $\mathrm{Dim}(T_{\psi(p)}\setR^n)$ as the cardinality of the basis of directional derivatives.

For future reference, we note that the $i$th coordinate for a vector $v \in T_{\psi(p)}\setR^n$ can be obtained by acting on the coordinate function along the basis vector $i \in \setR^n$. Directional derivatives along the basis of $\setR^n$, evaluated at $\psi(p)$, give a basis for $T_{\psi(p)}\setR^n$. 
\begin{equation}
\label{eqn:basis_TPpRn}
\begin{aligned}
&\text{for } v \in T_{\psi(p)}\setR^n \colon &
    & v^{(i)} = v \left(x \mapsto (x - \psi(p)) \cdot i \right) &\\
&& \text{and}\qquad
    & v = \sum_i v^{(i)} (i \cdot \restr{\nabla}{\psi(p)}) &
\end{aligned}
\end{equation}

In \fref{sec:manifolds} we complete this line of thought by using the same maps to transport a basis as well as component functions from $\setR^n$ to $T_pM$, and using them to chart the space of vector fields.

\todo{check all these commented items are somewhere}

\section{More on manifolds}\label{sec:manifolds}
The main technical gaps we encountered in formalising a theory of Lie groups stem from missing results about manifolds.
In particular, a Lie group can act on a manifold through the diffeomorphism group, 
while the Lie algebra of a Lie group is usually defined as a set of smooth vector fields. 
We formalise both concepts below, and use them in \fref{sec:lie}.

\subsection{The diffeomorphism group}\label{sec:diff}

The diffeomorphism group of a manifold $M$, denoted $\Diff(M)$, is the set of all diffeomorphisms $M \to M$, together with the group operation of function composition. In this section, we define $\Diff(M)$ as a set of partial functions and prove the group axioms hold for a suitable operation.

As we have seen in \fref{sec:manifolds-review},
Immler and Zhan define the space of differentiable real-valued functions on a manifold by using a predicate that fixes the behaviour of the function outside the carrier set to be the constant map $0$ \cite{immler2019}.
However, the underlying type of a manifold has, in general, no ``canonical'' element: we only require it to be equipped with a topology.
Thus manifolds in general may not have a zero element, and we cannot use \lisa|extensional0| to define $\Diff(M)$.

Instead, we use the partial function type \lisa|'a\<Rightarrow>'b option| (shortened to \lisa|'a\<rightharpoonup>'b|) from Isabelle/HOL's \lisa|Map| theory. 
This is similar in that behaviour outside the carrier set is mapped to a constant, \lisa|None|, but avoids additional constraints on the type of the manifold's points. 
Points in the source manifold are mapped to points in the destination manifold, wrapped in the value \lisa|Some|; they can be accessed using the function \lisa|the|:
\begin{equation}\label{eqn:the}
\text{\lisa|the x|} =
\begin{cases}
\text{\lisa|y|} & \text{if } \text{\lisa|x|} = \text{\lisa|Some y|,} \\
\text{\lisa|undefined|} & \text{if } \text{\lisa|x|} = \text{\lisa|None|.}
\end{cases}
\end{equation}
The domain of such a partial map $f$ (accessed as \lisa|dom f|) is the set of all values that are mapped to \lisa|Some|.
We consider it an additional advantage that \lisa|None| is never the result of applying a diffeomorphism to a point in the manifold (while $0$, for example, might be).

We call a diffeomorphism from $M$ to itself an automorphism, and provide both a locale definition for a (total) function which is automorphic on a manifold (similar to \lisa|diff_fun| in \fref{sec:manifolds-review}), as well as a predicate for the corresponding partial function.

\begin{lstlisting}
locale c_automorphism = diffeomorphism k charts charts f f'
  for k charts f f'

definition (in c_manifold) "automorphism f \<equiv>
    (\<exists>f'. c_automorphism k charts (\<lambda>x. the (f x)) f') \<and> dom f = carrier"
\end{lstlisting}

The diffeomorphism group is now defined exactly as we stated at the beginning of this section.
\begin{lstlisting}
definition "Diff \<equiv> {f. automorphism f}"
\end{lstlisting}
The main requirement for proving $\Diff(M)$ is a group (under composition \lisa|Diff_comp| that respects function domains and the identity map \lisa|Diff_id| restricted to the carrier) is to find a unique inverse for any diffeomorphism.
Uniqueness is guaranteed by our use of partial maps: this eliminates differing functions that are identical (only) on the manifold.\todo{is this sentence clear?}
Thus we show that \lisa|Diff| provides an interpretation of the locale \lisa|grp_on|, which defines a group, in the context of an arbitrary \lisa|smooth_manifold|. We present the locale \lisa|grp_on| with some definitions unfolded.

\enlargethispage{2\baselineskip} 
\begin{lstlisting}
locale grp_on =
  fixes S::"'a set" and pls::"'a\<Rightarrow>'a\<Rightarrow>'a" and z
  assumes add_assoc: "\<lbrakk>a \<in> S; b \<in> S; c \<in> S\<rbrakk> \<Longrightarrow> pls (pls a b) c = pls a (pls b c)"
      and add_mem: "\<lbrakk>a \<in> S; b \<in> S\<rbrakk> \<Longrightarrow> pls a b \<in> S"
      and add_zero: "a \<in> S \<Longrightarrow> pls z a = a \<and> pls a z = a"
      and zero_mem: "z \<in> S"
      and inv_ex: "\<forall>x\<in>S. \<exists>y\<in>S. pls x y = z \<and> pls y x = z"
\end{lstlisting}

\begin{lstlisting}
sublocale Diff_grp: grp_on Diff Diff_comp Diff_id
\end{lstlisting}



A mathematician would expect a definition as we have given for \lisa|grp_on|, with additive inversion and subtraction left implicit, and defined uniquely thanks to the group axioms.
However, the group axioms only imply that the inverse map $x \mapsto - x$ (unary minus is denoted \lisa|um| in Isabelle) is unique on the carrier set, \emph{not} the total underlying type.
Thus it is traditional to carry unary and binary minus (\lisa|um| and \lisa|mns|) operators as additional parameters in locale definitions, in case an interpretation later wants to fix behaviour outside the carrier set too.
We provide a second locale, \lisa|group_on_with|, for this setting, and show that any \lisa|grp_on| is also a \lisa|group_on_with| (though not uniquely), and vice versa.
Proofs about groups in general are then carried out in \lisa|group_on_with|.
\begin{lstlisting}
locale group_on_with = 
  fixes S::"'a set" and pls::"'a\<Rightarrow>'a\<Rightarrow>'a"
    and z and mns um
  assumes add_assoc: "\<lbrakk>a \<in> S; b \<in> S; c \<in> S\<rbrakk> \<Longrightarrow> pls (pls a b) c = pls a (pls b c)"
      and add_mem: "\<lbrakk>a \<in> S; b \<in> S\<rbrakk> \<Longrightarrow> pls a b \<in> S"
      and add_zero: "a \<in> S \<Longrightarrow> pls z a = a \<and> pls a z = a"
      and zero_mem: "z \<in> S"
      and left_minus: "a \<in> S \<Longrightarrow> pls (um a) a = z"
      and diff_conv_add_uminus: "\<lbrakk>a \<in> S; b \<in> S\<rbrakk> \<Longrightarrow> mns a b = pls a (um b)"
      and uminus_mem: "a \<in> S \<Longrightarrow> um a \<in> S"
\end{lstlisting}

We remark that we also formalised a variant of the diffeomorphism group (including proof of the group axioms) using the extensional function set \lisa|PiE| from the HOL-Library, which functions similarly to \lisa|extensional0|, but uses the \lisa|undefined| element instead of 0.
The \lisa|PiE| approach provides easier function application compared to \lisa|option|, and greater generality than \lisa|extensional0|.
However, it lacks the main convenience of both alternatives, \ie/ out-of-bounds behaviour useable in proofs.
It is possible to conclude, for example, $$f(x) = \texttt{None} \implies x \notin \mathrm{Dom}(f),$$ which has no analogue for $f(x) = \texttt{undefined}$.
We also felt it was useful to separate design decisions and logic by handling partiality on the type-level.
Finally, we note that the type \lisa|'a\<rightharpoonup>'b| is isomorphic\todo{poor word choice? equivalent? One could also argue, for some understanding of isomorphism, that this is true of all three options...} to a quotient type over pairs $(f,d)$, where $f$ is a function with domain $d$, and two such pairs are equivalent if they have the same domain, and the functions agree on this domain.
Thus it is perhaps the most mathematically intuitive solution.

\subsection{Smooth vector fields}\label{sec:svf}
We have reviewed how to generalise the notion of smooth functions, as well as that of tangent vectors, to manifolds (in \fref{sec:manifolds-review}).
Since we have an intuitive, Euclidean idea what it means for a vector-valued function to smoothly vary (\eg/ the velocity of a smooth trajectory, or the gradient of a smooth function), it is natural to augment our theory of manifolds with a corresponding definition.

\begin{definition}\label{def:vector-field}
Let $M$ be a manifold.
A rough \emph{vector field} $X$ is a function that assigns to every point $p \in M$ a tangent vector $X_p \in T_pM$.
\end{definition}
The corresponding Isabelle definition below carries over the \lisa|extensional0| clause seen in Immler and Zhan's definition of tangent spaces.
\begin{lstlisting}
definition "rough_vector_field X \<equiv> extensional0 carrier X \<and>
                                   (\<forall>p\<in>carrier. X p \<in> tangent_space p)"
\end{lstlisting}

Mathematical texts go on to show how the set of all tangent vectors can be given the structure of a manifold, and smoothness of vector fields is then just the smoothness of maps between manifolds.
Isabelle's type system (in combination with existing implementation choices \cite{immler2019}) bars that road to us.
We examine the problem, and propose a solution in \fref{sec:TM}. 

Since any tangent vector can be seen as a function $C^{\infty}(M) \to \setR$, we obtain a second interpretation of the vector field as a function $C^{\infty}(M) \to C^{\infty}(M)$, which maps a function $f$ to the function \hbox{$Xf: p \mapsto X_p(f)$}.
Of course, we now have to prove that the obtained function is indeed smooth, and, given our definition of \lisa|diff_fun_space|,
that it is 0 outside the manifold's carrier set (see \fref{sec:manifolds-review}).
We do so in \fref{sec:derivations}, and show smooth vector fields are precisely equivalent to derivations on the algebra $C^\infty(M)$. 

We define two type synonyms (\ie/ abbreviations for types) in Isabelle to ease our presentation of vector fields: the argument \lisa|X| in the definition above is of type \lisa|'a vector_field|, where \lisa|'a| is the underlying type of the manifold. A similar type, \lisa|'a tangent_bundle|, will be used to define smooth vector fields.
\begin{lstlisting}
type_synonym 'a vector_field = "'a \<Rightarrow> (('a\<Rightarrow>real)\<Rightarrow>real)"
type_synonym 'a tangent_bundle = "'a \<times> (('a\<Rightarrow>real)\<Rightarrow>real)"
\end{lstlisting}

\subsubsection{The tangent bundle}\label{sec:TM}

An experienced mathematician might simply define the tangent bundle as the fibre bundle formed by gluing together all the tangent spaces using the smooth transition functions between atlas charts.
For brevity's sake, we will not give a full account of fibre (and vector) bundles. 
Instead, to justify a formal definition of smooth vector fields below, we give the following, somewhat intuitive, idea of the tangent bundle.

\begin{definition}\label{def:TM}
The tangent bundle of $M$ is the disjoint union of all tangent spaces, equipped with the topology induced by the charts.
\[TM = \bigsqcup_{p \in M} T_pM = \{(p,v) \;|\; p \in M \land v \in T_pM \}  \]
In particular, a subset of $TM$ is open if it is a union of preimages of open sets under $\phi \times d\phi$, for any atlas chart $\phi$.
\todo{This is more succinctly said as ``topology generated by the basis of preimages ...'' - quite technical...}
\end{definition}

If we wanted to give this disjoint union the structure of a manifold, all we need to provide is a set of charts (and, implicitly, a topology).
In Isabelle, however, these charts must obey specific sort constraints: they must be functions from a type \lisa|'a| with a (second-countable Hausdorff) topology to a Euclidean type \lisa|'e|.
Since tangent vectors have type \lisa|('a\<Rightarrow>real)\<Rightarrow>real|, we are faced with two problems.
Firstly, a topology on this type is already defined (the product topology), and we cannot simply provide an alternative: a type can instantiate a class only once.%
\footnote{
    The product topology on function spaces is also called the topology of pointwise convergence \cite[App.~A]{lee2012}. We could have attempted to circumvent this problem by defining an alias or duplicate for the function type, but deemed this solution no better than our current implementation; it would also not resolve the second problem.
}
Secondly, our manifold is defined by its atlas, a \emph{set of charts}, not by the entire type \lisa|('a,'e)chart|.
In Isabelle's simple type system, we cannot parametrise an instantiation of a type to a class on our set of charts -- the resulting instance would be dependently typed.\todo{Talking out of my comfort zone here.}
Similarly, the topology used to formulate the axioms of charts must be a topology on the \emph{entire} underlying type, not just the disjoint union $TM$, and we cannot use the charts to induce a meaningful topology on the entire type.

As a shortcut through these issues, instead of using the charts to induce a topology on $TM$, we construct the functions that would be charts, and show that they are compatible (in the sense of a smooth manifold). In the notation of \fref{fig:tangent-space-maps} and \fref{eqn:basis_TPpRn},
the predicate \lisa|apply_chart_TM| (resp.\ \lisa|domain_TM|, \lisa|codomain_TM|, \lisa|inv_chart_TM|) takes a chart $\phi\colon U \to V$ and constructs the forward function (resp.\ domain, codomain, inverse function) of a local homeomorphism $\phi_{TM} : TM \to \setR^{2n}$ where\todo{maths below: are brackets in 2nd/3rd line needed around triple function composition?}
\begin{equation}\label{eqn:chart_TM}
\begin{aligned}
\phi_{TM} \;&: \{(p,v) \;|\; p \in U \land v \in T_pM\} \xrightarrow{\simeq} V \times \setR^n \\
\phi_{TM}(p,v) \;&= \left(
  \phi(p) \;,\;
    \sum_{i \in \mathrm{Basis}} (d\kappa \circ d\phi \circ d\iota^{-1} (v))^{(i)} i
\right) \\
\phi_{TM}^{-1}(x,u) \;&= \left(
  \phi^{-1}(x) \;,\;
    d\iota \circ d\phi^{-1} \circ d\kappa^{-1} (u \cdot \restr{\nabla}{x})
\right)
\end{aligned}
\end{equation}

The vector elements of the products above are really just the coordinate functions and directional derivatives for $T_{\psi(p)}\setR^n$ given in \fref{eqn:basis_TPpRn}, but mapped isomorphically to the correct tangent space.
One complication in proofs using these quantities is the dependence of inverses like $d\kappa^{-1}$ on the point $p$, which is no longer fixed in the context of vector fields.
\todo[inline]{This is in fact something I want to investigate further later. Why can we define the push-forward without explicit $p$-dependence, but both our inverses (obtained via \lisa|THE|) and I-Z's pull-backs have explicit dependence on the point the tangent space is defined at? I think there may be a relation to the theorem about the differential being a functor to the category of smooth bundles, but not anything similar in to the category of tangent spaces? Or is there a theorem, but with pointed-manifolds, and it just so happens this is easily omitted (accidentally) in only one direction?}

Full Isabelle definitions of these quantities (including the linear isomorphisms, \eg/ \lisa|d\<iota>| and \lisa|d\<kappa>\<inverse>|) are verbose, and postponed to \fref{sec:tangent-coord}, in favour of more concise discussion in this section. Despite the complexity of their definition, we provide lemmas that compute these quantities in the same form one would see in a textbook, when given valid arguments. In effect, using isomorphisms like $d\iota^{-1}$ in these definitions is just a way of keeping track of function restriction and extension. We give an example, \lisa|coordinate_vector_apply_in'|, of such a lemma for the vector part of $\phi_{TM}^{-1}$, \ie/ the coordinate vector along a Euclidean vector \lisa|b|, which reduces simply to a derivative (\lisa|frechet_derivative|) in a given chart -- just like the directional derivative would in a Euclidean space. The assumptions \lisa|\<psi> \<in> atlas| and \lisa|p \<in> domain \<psi>| are subsumed in the local context.
\begin{lstlisting}
lemma coordinate_vector_apply_in':
  assumes k: "k=\<infinity>" and f: "f\<in>diff_fun_space"
  shows "(d\<iota> \<circ> d\<psi>\<inverse> \<circ> d\<kappa>\<inverse> \<circ> directional_derivative \<infinity> (\<psi> p)) b f =
          frechet_derivative (f \<circ> inv_chart \<psi>) (at (\<psi> p)) b"
\end{lstlisting}

We also show that this construction indeed leads to compatibility between charts of $TM$ constructed from different atlas charts, as one would require of a smooth manifold. This is formalised in the lemma \lisa|atlas_TM|, which can be compared to the definition of \lisa|k-smooth_compat| for two charts in \fref{sec:manifolds-review}.
\begin{lstlisting}
lemma atlas_TM:
  assumes "c1 \<in> atlas" "c2 \<in> atlas"
  shows "smooth_on ((apply_chart_TM c1) ` (domain_TM c1 \<inter> domain_TM c2))
    ((apply_chart_TM c2) \<circ> (inv_chart_TM c1))"
\end{lstlisting}

Thus reassured, we define a function $X\colon M \to TM$ as smooth if its composition with the charts above gives a smooth map $M \to \setR^{2n}$, \ie/ if we can find a chart $c$ around any point $p \in M$ such that 
\begin{lstlisting}
smooth_on (codomain c) (apply_chart_TM c \<circ> X \<circ> inv_chart c)
\end{lstlisting}

This line is indeed the interesting part in the definition of \lisa|smooth_from_M_to_TM|: smoothness at a point $x$ in a chart $c$ is defined just as it is between manifolds (see \lisa|diff_fun| in \fref{sec:manifolds} for the simple case of real-valued functions), but with the charts of the destination manifold (which instantiates the locale \lisa|manifold|) replaced by the functions used to chart $TM$ (which we cannot interpret as a \lisa|manifold| locale). Such a charting function for $TM$ can be constructed from $c$ using \lisa|apply_chart_TM| as described above. The smoothness requirement listed above occurs at the end of the abbreviation \lisa|k_diff_from_M_to_TM_at_in| below.
\begin{lstlisting}
abbreviation (in c_manifold) k_diff_from_M_to_TM_at_in ::
  "enat \<Rightarrow> 'a \<Rightarrow> ('a,'b)chart \<Rightarrow> ('a \<Rightarrow> 'a tangent_bundle) \<Rightarrow> bool"
  where "k_diff_from_M_to_TM_at_in k' x c X  \<equiv>
    x \<in> domain c \<and>
    X ` domain c \<subseteq> domain_TM c \<and>
    k'-smooth_on (codomain c) (apply_chart_TM c \<circ> X \<circ> inv_chart c)"
\end{lstlisting}
The rest is straightforward: a function is smooth on a manifold if it is smooth in some chart at each point of the manifold. We specify an abbreviation for the case of $C^\infty$-manifolds.
\begin{lstlisting}
definition (in c_manifold) k_diff_from_M_to_TM (\<open>_-diff'_from'_M'_to'_TM\<close>)
  where diff_from_M_to_TM_def: "k'-diff_from_M_to_TM X \<equiv>
    \<forall>x. x \<in> carrier \<longrightarrow> (\<exists>c\<in>atlas. k_diff_from_M_to_TM_at_in k' x c X)"

abbreviation (in smooth_manifold) "smooth_from_M_to_TM \<equiv> k_diff_from_M_to_TM \<infinity>"
\end{lstlisting}

It is more convenient to work with vector fields as maps from $M$ to tangent vectors directly, as described in \fref{sec:svf}, rather than maps into the product type underlying $TM$.
Fortunately there is a simple correspondence: a vector field is a \emph{section} of $TM$, \ie/ its first component is the identity. Thus we can easily apply the smoothness requirement to a simpler type. 
\begin{lstlisting}
definition "smooth_vector_field X \<equiv> rough_vector_field X \<and>
                                    smooth_from_M_to_TM (\<lambda>p. (p, X p))"
\end{lstlisting}

Since \lisa|smooth_from_M_to_TM| mirrors the definition of smoothness between manifolds, directly filling in the charting functions we desire, it is easy to be convinced that it describes the wanted behaviour for vector fields.
Indeed, we prove that our definition leads to the same basic properties detailed in Lee's textbook \cite{lee2012}.
We outline some of these below and in \fref{sec:derivations}, leading up to the classic equivalence between derivations $C^\infty(M) \to C^\infty(M)$ and smooth vector fields $M \to TM$. 

A first important result is that we can fix a chart and work locally in coordinates without compromising our results.
The locale \lisa|smooth_vector_field_local| formalises the situation where, given an underlying manifold \lisa|charts|, a vector field \lisa|X| is smooth on the submanifold defined by the domain of a single chart \lisa|\<psi>|.
In this single-chart setting where coordinate representations are available, a vector field $X$ is smooth precisely if its component functions are smooth (\lisa|vector_field_smooth_local_iff|); and $X$ is smooth on an entire manifold if and only if it is smooth in any atlas chart (\lisa|smooth_vector_field_iff_local|, \cite[Prop.~8.1]{lee2012}).
Such lemmas are important to strengthen our definition: they rely on smooth compatibility of our charting functions.
(The assumption \lisa|vec_field_X| below can be read as ``$X$ is a rough vector field''.)\todo{no def is given for smooth-vector-field-local or vec-field-component}

\begin{lstlisting}
lemma vector_field_smooth_local_iff:
  assumes vec_field_X: "\<forall>p\<in>domain \<psi>. X p \<in> tangent_space p"
  shows "smooth_vector_field_local charts \<psi> X \<longleftrightarrow>
        (\<forall>i\<in>Basis. diff_fun_on (domain \<psi>) (vector_field_component X i))"
\end{lstlisting}
\begin{lstlisting}
lemma smooth_vector_field_iff_local:
  assumes "extensional0 carrier X"
  shows "smooth_vector_field X \<longleftrightarrow>
        (\<forall>c\<in>atlas. smooth_vector_field_local charts c X)"
\end{lstlisting}

\subsubsection{Coordinate functions for the tangent bundle}
\label{sec:tangent-coord}
In this section we give formal definitions of isomorphisms between tangent spaces as in \fref{fig:tangent-space-maps}, as well as the coordinate functions used to define charting maps for the tangent bundle in \fref{sec:TM}.
These technical details are interesting from the representational standpoint, but less relevant than the existence and properties of the charting functions described earlier. Skipping this section at a first reading will not detract from the rest of the paper.


Let $M$ be an $n$-dimensional smooth real manifold containing a point $p$. Let $U$ be an arbitrary submanifold of $M$, also containing $p$. Then there is a canonical inclusion $\iota \colon U \to M$, which is a smooth injection. By a standard result of differential geometry, this map has a pushforward $d\iota : T_pU \to T_pM$, which is in fact a linear isomorphism. As such, it has an inverse $d\iota^{-1}$; however, since $\iota$ is not a bijection, this inverse cannot be obtained as a pushforward of ``the inverse of $\iota$''.

In Isabelle, Immler and Zhan prove the standard result mentioned above in the locale \lisa|submanifold|, which fixes both a manifold and an open subset on which a submanifold structure is induced.
Once we fix an atlas chart $\psi$, whose domain is a submanifold of $M$, we can write down an inverse for $d\iota$ by using the definite selection operator (\lisa|the_inv_into|), leveraging locales to define convenient shorthand and notation like \lisa|d\<iota>\<inverse>|.
\begin{lstlisting}
locale c_manifold_local = c_manifold +
  fixes \<psi> assumes \<psi> [simp]: "\<psi> \<in> atlas"
begin

sublocale sub_\<psi>: submanifold charts k "domain \<psi>"
  by (unfold_locales, simp)

sublocale diffeo_\<psi>: diffeomorphism k
    "charts_submanifold (domain \<psi>)"
    "manifold_eucl.charts_submanifold (codomain \<psi>)"
    \<psi> "inv_chart \<psi>"
  <proof>

notation sub_\<psi>.inclusion.push_forward (\<open>d\<iota>\<close>)
\end{lstlisting}

\begin{lstlisting}
abbreviation differential_inclusion_inv_at (\<open>d\<iota>\<inverse>\<close>)
  where "d\<iota>\<inverse> p \<equiv> restrict0 (tangent_space p)
                 (the_inv_into (sub_\<psi>.sub.tangent_space p) d\<iota>)"
\end{lstlisting}
%
%
%
The functions $d\kappa$ and $d\kappa^{-1}$ are similarly defined, but on the Euclidean submanifold identified by the codomain of $\psi$ (see \fref{fig:tangent-space-maps}). The chart $\psi$ itself also has a push-forward: since charts are diffeomorphisms between their domain and codomain (\lisa|diffeo_\<psi>|), we obtain an inverse $d\psi^{-1} = d(\psi^{-1})$ in terms of the inverse chart.

Once we fix a point in $\mathrm{Dom}(\psi)$, we define notation for the various tangent spaces\todo{If pressed for space, can change $T_{\psi p}\psi U$ to $T_pU$ below and rm diffeo-PSI above} of \fref{fig:tangent-space-maps}.
We can even specialise our previous notation to be specific to the locally fixed point, so our notation is the same across the locales \lisa|c_manifold_local| and \lisa|c_manifold_point|.
This is syntactically ambiguous, but Isabelle's type inference correctly identifies the desired expression.
\begin{lstlisting}
locale c_manifold_point = c_manifold_local +
  fixes p assumes p [simp]: "p\<in>domain \<psi>"
begin
(*...*)
abbreviation "T\<^sub>\<psi>\<^sub>p\<psi>U \<equiv> diffeo_\<psi>.dest.tangent_space (\<psi> p)"
abbreviation dRestr2 (\<open>d\<iota>\<inverse>\<close>) where "dRestr2 \<equiv> differential_inclusion_inv_at p"
\end{lstlisting}

While it is possible to write down \lisa|the_inv_into f| for any function \lisa|f|, this has provable properties only when \lisa|f| is injective. Thus we use Immler and Zhan's results about \lisa|submanifold.inclusion.push_forward| to prove $d\iota^{-1}$ is also a linear isomorphism of tangent spaces.
\begin{lstlisting}
lemma bij_betw_d\<iota>_inv: "bij_betw d\<iota>\<inverse> T\<^sub>pM T\<^sub>pU"
lemma linear_on_d\<iota>_inv: "linear_on T\<^sub>pM T\<^sub>pU scaleR scaleR d\<iota>\<inverse>"
\end{lstlisting}
Similar results hold for the other tangent space isomorphisms.
One reason for restricting the main body of work to the case $k = \infty$ is that the identification of derivations in $T_{\psi(p)}E$ with directional derivatives in $\setR^n$ is one-to-one for smooth manifolds.
This allows us to invert it and obtain the coordinate maps below.\todo{check/cite}

Finally, in the locale \lisa|c_manifold_point|, we can define the coordinate system given in \fref{eqn:basis_TPpRn} (\lisa|coord_fun|), and use our isomorphisms to apply it to tangent vectors in $T_pM$ (\lisa|component_function|). Once we have coordinates for our tangent vectors, mapping them to a Euclidean space is a simple case of summing over the Euclidean basis (\lisa|tangent_chart_fun|).
\begin{lstlisting}
definition coord_fun where "coord_fun X i = X (\<lambda>x. (x - (\<psi> p)) \<bullet> i)"

definition component_function :: "(('a \<Rightarrow> real) \<Rightarrow> real) \<Rightarrow> 'b \<Rightarrow> real"
  where "component_function \<equiv> coord_fun \<circ> d\<kappa> \<circ> d\<psi> \<circ> d\<iota>\<inverse>"

definition tangent_chart_fun :: "(('a \<Rightarrow> real) \<Rightarrow> real) \<Rightarrow> 'b"
  where "tangent_chart_fun v \<equiv> \<Sum>i\<in>Basis. component_function v i *\<^sub>R i"
\end{lstlisting}
\todo[inline]{I imagine there's a simplification of tangent-chart-fun where the dot product and real scaling become a Dirac delta. I seem to remember working on that. Track down/try again? Did I abandon it because the domain algebra/function extensions look too messy?}

We obtain a similar transport of the basis of the charting Euclidean space (\lisa|coordinate_vector|), and prove that these quantities define a vector basis and the corresponding Euclidean representation (\lisa|coordinate_vector_representation|) on tangent spaces of smooth manifolds.

\begin{lstlisting}
definition coordinate_vector :: "'b \<Rightarrow> (('a \<Rightarrow> real) \<Rightarrow> real)"
  where "coordinate_vector = d\<iota> \<circ> d\<psi>\<inverse> \<circ> d\<kappa>\<inverse> \<circ> (directional_derivative k (\<psi> p))"
\end{lstlisting}

\begin{lstlisting}
lemma coordinate_vector_representation:
  assumes "v \<in> T\<^sub>pM" and "k=\<infinity>"
  shows "v = (\<Sum>i\<in>Basis. (component_function v i) *\<^sub>R (coordinate_vector i))"
\end{lstlisting}

We are now very close to constructing our candidate charts for the tangent bundle. All we need is to use these definitions for a variable parameter $p$. Isabelle allows us to invoke definitions inside locales in qualified syntax, \eg/ \lisa|c_manifold_point.coord_fun|, and provide any parameters of the locale as additional input. Finally, the charting functions given in \fref{eqn:chart_TM} can be translated to Isabelle as follows, constructing a quadruple similar to the defining representation of a \lisa|chart| (cf. \fref{sec:manifolds-review}).
\begin{lstlisting}
definition apply_chart_TM :: "('a,'b)chart \<Rightarrow> 'a tangent_bundle \<Rightarrow> 'b \<times> 'b"
  where "apply_chart_TM c \<equiv>
    \<lambda>(p, v). (
      c p,
      c_manifold_point.tangent_chart_fun charts \<infinity> c p v
    )"

definition inv_chart_TM :: "('a,'b)chart \<Rightarrow> ('b\<times>'b) \<Rightarrow> 'a\<times>(('a\<Rightarrow>real) \<Rightarrow> real)"
  where "inv_chart_TM c \<equiv>
    \<lambda>(p::'b, v::'b). (
      inv_chart c p,
      c_manifold_point.coordinate_vector charts \<infinity> c (inv_chart c p) v
    )"
\end{lstlisting}
\begin{lstlisting}
definition domain_TM :: "('a,'b) chart \<Rightarrow> ('a \<times> (('a \<Rightarrow> real) \<Rightarrow> real)) set"
  where "domain_TM c \<equiv> {(p, v). p \<in> domain c \<and> v \<in> tangent_space p}"

definition codomain_TM :: "('a,'b) chart \<Rightarrow> ('b\<times>'b) set"
  where "codomain_TM c \<equiv> {(p, v). p \<in> codomain c}"
\end{lstlisting}


\subsubsection{Smooth vector fields as derivations}\label{sec:derivations}

Since vector fields are functions of two variables of different type, it makes sense to develop their theory not only from the viewpoint of maps $M \to TM$, but also under the lens of self-mappings of the space of smooth functions, $C^\infty(M) \to C^\infty(M)$.
In particular, Lee states the following result, summarising how vector fields are determined by their action on function spaces, define closed operations on the space $C^\infty(M)$, and capturing a similar local condition to the one exhibited in \fref{sec:TM}.

\begin{theorem}[{\cite[Prop.~8.14]{lee2012}}]\label{thm:vector-field-smooth-iff}
Let $M$ be a smooth manifold [...\footnote{We have removed references to boundaries: like Immler and Zhan \cite{immler2019}, we only consider manifolds without boundaries.}], and let $X : M \to TM$ be a rough vector field. The following are equivalent:
\begin{enumerate}
\item[(a)] $X$ is smooth.
\item[(b)] For every $f \in C^\infty(M)$, the function $X f$ is smooth on $M$.
\item[(c)]\label{item:vf-smooth-iff-subset} For every open subset $U \subseteq M$ and every $f \in C^\infty(U)$, the function $X f$ is smooth on $U$.
\end{enumerate}
\end{theorem}

While closure over $C^\infty(M)$ is the most important part of this theorem with a view to the ultimate result of this section, proving Property~\textit{(c)} requires some refinement of the definitions of vector fields and their application to functions.
As explained in \ref{sec:manifolds-review}, the tangent spaces $T_pM$ and $T_pU$ are isomorphic, but not identical; similarly for the smooth function spaces $C^\infty(M)$ and $C^\infty(U)$.
Furthermore, the operation of a vector field is only non-zero on smooth functions from the correct space, \ie/ a vector field cannot act on both a function and a non-trivial restriction of that function.
The innocuous restriction-up-to-isomorphism (reproduced from \cite[p.~180]{lee2012})
\begin{equation}\label{eqn:restrict1}
\restr{(Xf)}{U} = X(\restr f U)
\end{equation}
is more elaborate in Isabelle, because meaningfully applying a vector field that is smooth on $M$ to elements of $C^\infty(U)$ requires explicit application of the tangent space isomorphism $T_pM \simeq T_pU$. In the notation of \fref{fig:tangent-space-maps}, for $f \in C^\infty(M)$ and 
$X : M \to TM$, this equality should read
\begin{equation*}\label{eqn:restrict2}
\restr{(Xf)}{U} = (d\iota^{-1} \circ X) (\restr f U) \;.
\end{equation*}

We can expect the reader to infer whether to treat $X$ as a function on $M$ (as in \fref{def:vector-field}) or on $C^{\infty}(M)$ (as in $X f$ above). In Isabelle, we need to make it explicit when we want a vector field to apply to functions. We define notation for this (\lisa|\<hungarumlaut>|), and for restriction of vector fields as above (\lisa|\<restriction>|), using which we can write and prove \fref{eqn:restrict1} in Isabelle.\todo{can rm some assms? is the f assumption handled by extensionality?}
\begin{lstlisting}
lemma (in smooth_manifold) vec_field_apply_fun_in_restrict0':
    "restrict0 U (X\<hungarumlaut>f) = X\<restriction>U \<hungarumlaut> (restrict0 U f)"
  if "open U" "U \<subseteq> carrier" "f \<in> diff_fun_space" "rough_vector_field X" for U X f
\end{lstlisting}
Using this notation, we state and prove \fref{thm:vector-field-smooth-iff}.
\begin{lstlisting}
lemma vector_field_smooth_iff':
  fixes C_inf
  defines "\<And>U. C_inf U \<equiv> c_manifold.diff_fun_space (charts_submanifold U) \<infinity>"
  assumes X: "rough_vector_field X"
  shows "smooth_vector_field X \<longleftrightarrow> (\<forall>f\<in>diff_fun_space. (X \<hungarumlaut> f) \<in> diff_fun_space)"
    and "smooth_vector_field X \<longleftrightarrow> (\<forall>U f. open U \<and> U \<subseteq> carrier \<and> f \<in> C_inf U \<longrightarrow>
                                          diff_fun_on U (X\<restriction>U \<hungarumlaut> f))"
\end{lstlisting}

This states that vector fields act on smooth functions in a way that preserves smoothness.
The Isabelle variant above also preserves extensionality over the \lisa|diff_fun_space|.
This is reminiscent of tangent vectors acting on smooth functions at a point, and motivates the definition of a \emph{derivation on a manifold} (as opposed to a \emph{derivation at a point}) as a linear (\lisa|real_linear_on|) function $C^\infty(M) \to C^\infty(M)$ which obeys the Leibniz rule.
The final line in the listing of the definition \lisa|is_derivation_on|:
\begin{lstlisting}
definition is_derivation_on :: "(('a\<Rightarrow>real) \<Rightarrow> ('a\<Rightarrow>real)) \<Rightarrow> bool"
  where "is_derivation_on D \<equiv>
    real_linear_on diff_fun_space diff_fun_space D \<and>
    (\<forall>f\<in>diff_fun_space. \<forall>g\<in>diff_fun_space. D (f*g) = f*(D g) + g*(D f)) \<and>
    D ` diff_fun_space \<subseteq> diff_fun_space"
\end{lstlisting}
just adds an intuitive condition on the codomain -- this is missing in \lisa|linear_on|.

Using \fref{thm:vector-field-smooth-iff}, we show that smooth vector fields are in one-to-one correspondence with derivations, provided both the derivations, like our vector fields, are understood to be \lisa|extensional0| over the carrier and smooth function sets. 

\enlargethispage{-2\baselineskip} 
\begin{lstlisting}
theorem smooth_vector_field_iff_derivation:
  fixes extensional_derivation
  defines "\<And>D. extensional_derivation D \<equiv> is_derivation_on D \<and>
    extensional0 carrier (\<lambda>p f. D f p) \<and> extensional0 diff_fun_space D"
  shows "smooth_vector_field X \<Longrightarrow> extensional_derivation (\<lambda>f. X \<hungarumlaut> f)"
    and "extensional_derivation D \<Longrightarrow> smooth_vector_field (\<lambda>p f. D f p)"
\end{lstlisting}

\section{Lie groups}\label{sec:lie}

A Lie group $G$ is a smooth manifold with a smooth group operation and smooth inverses.
Since the group operation is binary, this differentiability requirement applies to the curried group operation as a map $(\cdot) : G \times G \to G$ defined on the product manifold.
This is captured as the abbreviation \lisa|diff_on_product_manifold|.\footnote{A definition would have needed to be manually unfolded to relate to the defining concepts of differentiability and product. An abbreviation results in fewer simplifier rules while increasing readability.}
We represent the Lie group as a locale, fixing parameters for the group operation \lisa|times| (which also defines inverses \lisa|invs|; see below) and identity \lisa|one| as well as the manifold-defining set of \lisa|charts|.

\begin{lstlisting}
abbreviation "diff_on_product_manifold charts binary_op \<equiv>
  diff \<infinity> (prod_charts charts charts) charts (\<lambda>(a,b). binary_op a b)"
\end{lstlisting}

\enlargethispage{-1\baselineskip}
\begin{lstlisting}
locale lie_grp = c_manifold charts \<infinity> + grp_on carrier times one
  for charts :: "('a::{t2_space, second_countable_topology},
                  'e::euclidean_space) chart set"
    and times one +
  assumes smooth_mult: "diff_on_product_manifold charts times"
    and smooth_inv: "diff \<infinity> charts charts invs"
\end{lstlisting}

The type class constraints for \lisa|'a| and \lisa|'e| above mirror the topological requirements of manifolds in \autoref{sec:manifolds}.
We can then derive some initial lemmas about the group operations, such as the following smoothness result about multiplication as a map $G \to G$:
\begin{lstlisting}
lemma diff_times:
  assumes "x \<in> carrier"
  shows "diff \<infinity> charts charts (\<lambda>y. times x y)"
\end{lstlisting}
This relies on extending the standard library of Isabelle with a fundamental theorem about product manifolds: the \lisa|Pair| constructor maps any pair of smooth functions to a smooth function on the corresponding product manifold.

Exactly like the pair of locales \lisa|grp_on| and \lisa|group_on_with| in \fref{sec:diff}, our formalisation contains a second locale \lisa|lie_group|, which carries explicit parameters for inverse and division, and in which we carry out our proofs in general. Again, any \lisa|lie_grp| is also a \lisa|lie_group| (though not uniquely), and vice versa (uniquely).


\subsection{Instances of Lie groups}\label{sec:lie-models}

As a simple check of consistency of \lisa|lie_group|, we show that Euclidean spaces form Lie groups under vector addition. Both operations \lisa|+| and \lisa|-| (and unary minus) are axiomatic constants on the class of Euclidean spaces.
The Euclidean manifold is defined by a single chart, the identity map, which is the sole member of \lisa|charts_eucl|.
\begin{lstlisting}
interpretation lie_group_eucl: lie_group charts_eucl "(+)" 0 "(-)" uminus
\end{lstlisting}

As a more involved model, we show the general linear group $\GL$ is a Lie group.
This required a surprising amount of effort and machinery, including two theories about different matrix types, the transfer package (a generic tool for specifying type equivalence and transferring results between equivalent types) \cite{huffman2013}, and a number of proofs about smoothness of matrix constants.
The main missing ingredient for this proof is smoothness of the determinant: it implies both that $\GL$ is a manifold and that matrix inversion is smooth.
\begin{lstlisting}
lemma smooth_on_det:
  fixes s::"(('a::real_normed_field^'n)^'n) set"
  assumes "open s"
  shows "smooth_on s det"
\end{lstlisting}
Indeed, continuity of the determinant is sufficient to show that
$\GL = {\det}^{-1} \left(\setR \setminus \{0\}\right)$
is open, and thus a submanifold.
The inverse of a matrix can be written in terms of its adjugate and its determinant, as shown by Adelsberger et al.\ in the AFP's Cayley-Hamilton theory \cite{Cayley_Hamilton-AFP}.
Adelsberger et al.\ prove this matrix inverse formula on a type of square matrices whose representation differs from the matrix type of Isabelle/HOL's analysis library.%
\footnote{A square matrix has type \lisa|'a^'n^'n|, where \lisa|'n| is a finite type whose size determines the dimension of the matrix with elements in \lisa|'a|. This type of matrix is based on work by Harrison in HOL Light \cite{harrison2013}.}
However, we expect that matrix theorems exist for both representations: while they are different, they formalise equivalent mathematical concepts.
In this kind of situation, the transfer package allows us to avoid proof duplication by transporting theorems between equivalent types \cite{huffman2013,kappelmann2023}.\todo{We're not using Kappelmann's formulation, can I still cite him?}

\begin{lstlisting}
lemma inverse_adjugate_det:
  fixes A::"'a::field^'n^'n"
  assumes "invertible A"
  shows "matrix_inv A =  (1 / (det A)) *\<^sub>s (adjugate A)"
\end{lstlisting}

Without going into technical details, transfer rests on the application of transfer rules, which specify an equivalence between specific constants on two types.
Writing down a transfer rule generates a proof obligation to ensure this equivalence is true; upon fulfilment of that obligation by the user, the transfer rule is registered for use by the transfer algorithm.
Once required transfer rules for the constants appearing in a theorem proven on one type are present, the package automates a proof of the equivalent statement on the equivalent type.
In our case, the required transfer rules link representations of constants such as the adjugate of a matrix, the identity matrix, addition, and matrix powers.
The advantage is that transfer rules are reusable across many theorems, may be easier to prove than involved theorems, and that any constants appearing in the proof, but not the theorem statement, do not require separate transfer rules.
For example, while our target is the matrix inverse, we also transfer the Cayley-Hamilton theorem itself, with very little additional effort: as a rough measure, our entire transfer setup takes up 216 lines of proof script, compared to 842 for the AFP entry of Adelsberger et al.. Other work involving transfer between matrix types is summarised by Divas\'{o}n and Thiemann \cite{divason2022}.

We then combine the transferred matrix inverse formula with our proof of the smoothness of the determinant: since the adjugate matrix elements are polynomials (thus smooth),  we can conclude that matrix groups have smooth inverses.


\begin{lstlisting}[commentstyle=]
theorem real_GL_Lie_grp: "lie_grp real_charts_GL (**) (mat 1)"
\end{lstlisting}
Again, the only chart required is the identity, but restricted to the set \lisa|GL| of invertible matrices.


\subsection{Lie groups acting on manifolds}\label{sec:action}
Given a manifold $M$ and its diffeomorphism group $\Diff(M)$, the action of a Lie group $G$ on the manifold $M$ is a group homomorphism $\rho: G \to \Diff(M),\; g \mapsto \rho_g$ such that the map $(g,m) \mapsto \rho_g (m)$ is smooth.
We formalise this definition in the locale \lisa|lie_group_action|, abbreviating the smoothness axiom for $(g,m) \mapsto \rho_g (m)$ using \lisa|diff_action_map| (\cf/~the abbreviation \lisa|diff_on_product_manifold| in the definition of \lisa|lie_grp| in \fref{sec:lie}).
To obtain an element of the manifold $M$ after applying a partial map \lisa|\<rho> g \<in> Diff|, we unwrap the result using \lisa|the|, which maps \lisa|None| to \lisa|undefined| (see \fref{eqn:the}).
Our locale axioms ensure output is always well-defined for inputs that lie on the manifold.
We omit our definition of homomorphisms, \lisa|group_hom_betw|, but refer the reader to definitions of similar morphisms in \fref{sec:locales}.
For legibility of the definition below, we omit some locale parameters that are the same as elsewhere in this paper (\eg/ the definition of \lisa|lie_group| in \fref{sec:lie} or \lisa|Diff_grp| in \fref{sec:diff}).

\begin{lstlisting}
abbreviation (input) "diff_action_map g_charts m_charts action \<equiv>
  diff \<infinity> (c_manifold_prod.prod_charts g_charts m_charts) m_charts action"
\end{lstlisting}

\begin{lstlisting}
locale lie_group_action =
    lie_group charts (*...*) +
    M: c_manifold m_charts k +
    A: group_hom_betw carrier M.Diff (*...*) \<rho>
  for charts and m_charts and (*...*) \<rho> :: "'a \<Rightarrow> ('b\<rightharpoonup>'b)" +
  assumes "diff_action_map charts m_charts (\<lambda>(g,m). the ((\<rho> g) m))"
\end{lstlisting}

Every Lie group can act on itself through the group operation. This leads to the left, right, and adjoint self-actions, of which we only present the first here. We use the operator~\lisa!`|! below, which restricts a function \lisa|'a\<Rightarrow>'b| to a partial map \lisa|'a\<rightharpoonup>'b| on a given set.

\begin{lstlisting}
abbreviation "left_action g \<equiv> (times g) `| carrier "
\end{lstlisting}

Showing the left action $\mathcal L$ is indeed an action requires the lemma
\lisa|diff_tms| of \fref{sec:lie},
and implies that any \lisa|lie_group| interprets the locale \lisa|lie_group_action|, where the manifold acted upon is the Lie group itself as a $C^\infty$-manifold.

\begin{lstlisting}
sublocale left_action: lie_group_action
    charts times one (*...*)      (* the acting Lie group *)
    charts \<infinity>                 (* the manifold acted upon *)
    left_action               (* the action map *)
\end{lstlisting}

\subsection{Lie algebras}\label{sec:lie-algebras}

One of the central theorems about Lie groups is that every Lie group can be associated with a Lie algebra, even though, a priori, these are completely different structures.
\enlargethispage{-2\baselineskip} 
\begin{definition}[Lie algebra]
An algebra $(\mathfrak g, +, [\_,\_])$ is called a \emph{Lie algebra} if its bilinear form $[\_,\_]$ (the \emph{Lie bracket}) is alternating (antisymmetric) and obeys the Jacobi identity: 
\begin{equation}
    \forall x,y,z \in \mathfrak g: \qquad [x, [y, z]] + [y, [z, x]] + [z, [x, y]] = 0 
\end{equation}
\end{definition}
In Isabelle:

\begin{lstlisting}
locale lie_algebra =
   algebra_on \<gg> scale lie_bracket +
   alternating_bilinear_on \<gg> scale lie_bracket
 for \<gg> and scale :: "'a::field \<Rightarrow> 'b::ab_group_add \<Rightarrow> 'b"
   and lie_bracket :: "'b \<Rightarrow> 'b \<Rightarrow> 'b" +
 assumes jacobi: "\<lbrakk>x\<in>\<gg>; y\<in>\<gg>; z\<in>\<gg>\<rbrakk> \<Longrightarrow> 0 = [x;[y;z]] + [y;[z;x]] + [z;[x;y]]"
\end{lstlisting}
Given a Lie algebra $\mathfrak g$, a Lie subalgebra is a vector subspace $\mathfrak h$ that is closed under the Lie bracket. Unsurprisingly, $\mathfrak h$ is also a Lie algebra under the operations of $\mathfrak g$. The prefix\todo{word choice?} \lisa|m1| collects facts about $\mathfrak g$ as a module, so that \lisa|m1.subspace| means ``is a vector subspace of $\mathfrak g$''.
This naming of the source module (or vector space) as \lisa|m1| is inherited from locales about maps between two modules: a near-identical example, as far as reference to source and destination modules is concerned, is \lisa|module_hom_on| in \fref{sec:locales}.
A \lisa|subspace| is a subset of a module (or vector space) that is closed under addition and scaling, and contains the zero element (definition adapted for presentation from Isabelle's standard distribution).
\begin{lstlisting}
definition (in module_on) subspace :: "'b set \<Rightarrow> bool" where
  "subspace T \<longleftrightarrow> 0 \<in> T \<and>
    (\<forall>x\<in>T. \<forall>y\<in>T. x + y \<in> T) \<and>
    (\<forall>c. \<forall>x\<in>T. c *s x \<in> T)"

lemma lie_subalgebra:
  assumes closed: "\<And>x y. x \<in> \<hh> \<Longrightarrow> y \<in> \<hh> \<Longrightarrow> lie_bracket x y \<in> \<hh>"
    and subspace: "\<hh> \<subseteq> \<gg>" "m1.subspace \<hh>"
  shows "lie_algebra \<hh> scale lie_bracket"
\end{lstlisting}

This section is dedicated to constructing a Lie algebra of smooth vector fields; in particular, the left-invariant vector fields of a Lie group $G$ form \emph{the} Lie algebra $\mathrm{Lie}(G)$.
A vector field $X$ on a Lie group $G$ is left-invariant if the left action $\mathcal L_g$ of any element $g$ ``commutes'' with the vector field:
\begin{lstlisting}
definition (in c_manifold) vector_field_invariant_under (infix \<open>invariant'_under\<close> 80)
  where "X invariant_under F  \<equiv>  \<forall>p\<in>carrier. \<forall>f\<in>diff_fun_space.
    X (F p) f = (diff.push_forward k charts charts F) (X p) f"

abbreviation (in lie_group) "L_invariant X \<equiv> \<forall>p\<in>carrier. X invariant_under (\<L> p)"
\end{lstlisting}



This section brings together much of the preceding formalisation: left-invariance relies on knowledge of actions of Lie groups (\fref{sec:action}), which in turn are defined using the diffeomorphism group (\fref{sec:diff}).
Smoothness of vector fields occupied us in \fref{sec:svf}, and we will directly use their relation to derivations in our proofs below.
We will first show the space of smooth vector fields, denoted $\svf$, is a Lie algebra, and proceed to prove the left-invariant smooth vector fields, denoted $\svf_\mathcal{L}$, are a Lie subalgebra.
The operation of the Lie algebra $\svf$ is called the \emph{Lie bracket of smooth vector fields}. It is comparable to a commutator, and denoted by square brackets.
\begin{lstlisting}
lemma lie_bracket_def: "[X;Y] p f = X p (Y\<hungarumlaut>f) - Y p (X\<hungarumlaut>f)"
\end{lstlisting}

The Lie bracket is clearly antisymmetric, and it preserves extensionality of our \lisa|rough_vector_field|s. Our proof that it satisfies the Leibniz rule, \ie/ preserves derivations, is remarkably close to the textbook account of Lee \cite[p.~186]{lee2012}. We partly reproduce it below, cutting standard setup and method applications (marked resp.\ with \lisa|(*...*)| and \lisa|<proof>|). In particular, the \emph{only} statements cut are local simp rules and interpretations that make available linearity results for $X$ and $Y$ as maps on the smooth function space. Similarly, all method applications omitted are single applications of simple automatic Isabelle commands (\lisa|simp|, \lisa|presburger|, and \lisa|auto|), and discharge their goal.

\begin{lstlisting}
lemma product_rule_lie_bracket:
  assumes X: "smooth_vector_field X" and Y: "smooth_vector_field Y"
    and diff_funs: "f\<in>diff_fun_space" "g\<in>diff_fun_space"
  shows "[X;Y] \<hungarumlaut> (f * g) = f * [X;Y] \<hungarumlaut> g + g * [X;Y] \<hungarumlaut> f"
proof -
  (*...*)
  have "([X;Y] \<hungarumlaut> (f*g)) = X \<hungarumlaut> (Y\<hungarumlaut>(f*g)) - Y \<hungarumlaut> (X\<hungarumlaut>(f*g))" <proof>
  also have "\<dots> = X \<hungarumlaut> (f*Y\<hungarumlaut>g + g*Y\<hungarumlaut>f) - Y \<hungarumlaut> (f*X\<hungarumlaut>g + g*X\<hungarumlaut>f)" <proof>
  also have vf_linear: "\<dots> = X \<hungarumlaut> (f*Y\<hungarumlaut>g) + X \<hungarumlaut> (g*Y\<hungarumlaut>f) - Y \<hungarumlaut> (f*X\<hungarumlaut>g) - Y \<hungarumlaut> (g*X\<hungarumlaut>f)" <proof>
  also have "\<dots> = (f * X\<hungarumlaut>(Y\<hungarumlaut>g)) + (g * X\<hungarumlaut>(Y\<hungarumlaut>f)) - (f * Y\<hungarumlaut>(X\<hungarumlaut>g)) - (g * Y\<hungarumlaut>(X\<hungarumlaut>f))" <proof>
  finally show ?thesis <proof>
qed
\end{lstlisting}

We list this proof sketch to showcase how despite the troubles we faced when defining differential structures, such as smooth vector fields, simple algebraic proofs using them remain simple in Isabelle, and if one ignores the need to bring linearity and simp rules into scope, run very close to the pen and paper reasoning.
For completeness, we note that the step labelled \lisa|vf_linear| is not in Lee's book: Isabelle needs a clue to invoke linearity properties of \lisa|X| and \lisa|Y| on the function space. In exchange, Lee's final two simplification steps are easily performed together by Isabelle's simplifier.

{ 
}

The proof listed above is the main part of showing that the Lie bracket is closed on $\svf$: since $[X,Y]$ obeys the Leibniz rule, it is a derivation, so it is enough to invoke \lisa|is_derivation_on| (\fref{sec:derivations}).
Because our definitions use \lisa|extensional0|, $\svf$ is a vector space under the standard operations on the underlying type, so no definition is needed for addition \lisa|(+)| and real scaling \lisa[commentstyle=]|(*\<^sub>R)|.
Proving the Jacobi identity for the Lie bracket follows a similar profile to the proof relating to the Leibniz rule: we bring local interpretations and simp rules into scope, and closely follow a pen and paper outline \cite[p.~188]{lee2012}.

\begin{lstlisting}
lemma lie_bracket_jacobi: "[X; [Y;Z]] + [Y;[Z;X]] + [Z;[X;Y]] = 0"
   if "smooth_vector_field X" "smooth_vector_field Y" "smooth_vector_field Z"
\end{lstlisting}

And so we conclude smooth vector fields do indeed form a Lie algebra. We point out again that no domain assumptions are necessary on the elements $p\in M$ and the functions $f \in C^\infty(M)$ the bracketed vector fields apply to. Inside the proof context, we assume all inputs are well-chosen for the main proof, and show extensionality trivially implies the desired conclusion otherwise.

So far, we've only needed a generic manifold, and made no mention of multiplication on the manifold. If we do consider Lie groups instead, we can refine our Lie algebra to $\svf_{\mathcal L}$, the left-invariant (smooth) vector fields. The following lemma uses \lisa|lie_subalgebra| from the beginning of this section.
\begin{lstlisting}[commentstyle=,moreliterate={\\<XX>\\<^sub>\\<L>}{{$\mathfrak{X}_{\mathcal{L}}$}}2]
lemma lie_algebra_of_left_invariant_svf:
  fixes \<XX>\<^sub>\<L> defines "\<XX>\<^sub>\<L> \<equiv> {X. smooth_vector_field X \<and> L_invariant X}"
  shows "lie_algebra \<XX>\<^sub>\<L> (*\<^sub>R) (\<lambda>X Y. [X;Y])"
\end{lstlisting}
Conventionally, $\mathfrak X_\mathcal{L}$ is referred to as \emph{the} Lie algebra of the Lie group it is constructed on.

This result is the foundation of the Lie group-algebra correspondences that Lie theory revolves around.
Obtaining the Lie algebra in Isabelle/HOL is both a strong demonstration of the expressive capacity of Isabelle's simple type theory, and an exposition of many of the technologies that aid interactive proof.
Examples of these include sledgehammer, which often proves intermediate facts for us, and leaves us to follow the mathematical path through a proof instead of getting absorbed in lowest-level details; transfer, which gives us the some flexibility in the use of existing results independently from representational decisions made by their authors; and locales, which alleviate reasoning with set-based structures by providing a persistent, extensible, and hierarchical way of packaging common assumptions.\todo{I dislike this paragraph. Strong criticism welcome.}

\section{Comparison with existing developments}\label{sec:comparison}
We have spent significant effort navigating the restrictions of simple type theory during this development, and a natural reaction to, for example, the problems of \fref{sec:TM} is to wonder if a dependent type would not be a better fit for a formalisation of manifolds.
We therefore briefly compare our work with two similar formalisations in Lean \cite{demoura2015} using its intensional type theory (ITT):
Cavalleri and Bordg's development of Lie groups and vector bundles \cite{bordg2021}, and a separate theory of Lie algebras by Nash \cite{nash2022}.

Let us compare content first.
The MathLib theory of Lie algebras is significantly more developed than our contribution here, containing both an algebraic development of abstract Lie algebras, and constructions of the classical and exceptional Lie algebras.
We only provide an abstract definition, and are then interested in the link with manifolds and Lie groups.
Both our Isabelle/HOL theory and MathLib contain a Lie algebra associated with a manifold. 
The main difference here is that we also obtain an equivalence between derivations on $C^\infty(M)$ and smooth vector fields, and can therefore define left-invariance as a property of vector fields.
We could not find the analogous equivalence result in MathLib, even though there is a type class for tangent bundles.

This leads us to the most interesting contrast: the difficulties encountered by both approaches when working with smooth vector fields and/or derivations.
Our need for a new definition of smoothness (\lisa|smooth_from_M_to_TM|, \fref{sec:TM}) is mirrored with a new definition of \emph{heterogeneous} differentials in Lean.
While we were unable to produce the topology required by the sort constraints on the existing class of charts,
ITT's subtle handling of equality rendered insufficient Lean's existing definition of the differential of a smooth function.
Thus, as we define a variant of smoothness for vector fields, Cavalleri and Bordg define a variant of the differential for derivations.
There is potential for both of these workarounds to become more fundamental: in Isabelle one could parametrise the \lisa|manifold| locale with an underlying topology, and in Lean the heterogeneous definition with suitable type parameters specifies a homogeneous one.\todo{very much check with Ramon/Paul.} 

Our work on formalising Lie groups and algebras follows directions set by several recent locale-based formalisations of advanced mathematics.
The largest\todo{word} of these may be Bordg et al.'s formalisation of complex definitions of algebraic geometry in Isabelle/HOL \cite{bordg2022}, in response to a challenge raised by the Lean formalisation of Schemes \cite{buzzard2022}.
Both this impressive achievement, and our own contribution, follow a locale-centric approach best described (as mentioned previously) by Ballarin's exploration of textbook algebra \cite{ballarin2020}.
We remark that the locales we define for algebraic structures on sets are similar to those of Ballarin and Bordg et al., and if required, it is therefore easy to use results from these theories and ours side-by-side in future theories.

The importance of the Types-to-Sets (TTS) framework, employed to great effect by Immler and Zhan, and enhanced by Milehins \cite{immler2019,milehins2022}, must be recognised as well.
The transfer package itself, which underlies TTS, has recently been generalised by Kappelmann \cite{kappelmann2023}.
We use both of these libraries for their (semi-)automatically obtained results about vector spaces with explicit carrier sets, generated by TTS from type-based theories.
Without automation for such transfers, efforts like the present work would require prohibitive amounts of proof about locale-based alternatives to already well-supported type structures; in fact, Immler and Zhan's work using TTS is essential to the success of their formalisation, which this current work extends.
As an aside: Bordg et al.\ do not require TTS or its supporting framework of local type definitions \cite{kuncar2016}, but their effort is significant, while remaining tightly focussed on obtaining a definition of Groethendieck's Schemes.

\section{Conclusion and Outlook}\label{sec:conclusion}
Our formalisation of the Lie algebra of a Lie group includes varied content in differential geometry and linear algebra, reinforces some lessons in locale-based representation of mathematical hierarchies, and exemplifies several approaches to handling partial functions.
Regarding the second point in particular, we are considering a future extension of the theory with locale-based topology, in which the tangent bundle interprets the locale of smooth manifolds.
Content-wise, logical extensions include results on Lie groups, such as representation theory of the classical groups\todo{I include this because we're going to do (some) of it - at least formalise the classical groups up to $\mathrm{SO}(n)$ to get the Poincare group.};
a Lie algebra isomorphism between $T_1G$ and $\svf_\mathcal{L}$; 
and enhanced calculational tools, such as evaluation along curves and introduction of ODEs into the theory.
All of our definitions already allow for calculation in coordinates, for which we provide some simplification lemmas.

Locales prove, time and again, to be an invaluable foundation for advanced mathematics in Isabelle/HOL, enabling easier reasoning about substructures, and simultaneous interpretation of multiple similar structures on a single carrier set.
While TTS, particularly with recent extensions, is helpful, the authors hope that automation regarding transfer along isomorphisms can be improved, leading to easier reasoning ``up to isomorphism''.
In the current work, while we define quantities like coordinate functions for vector fields by explicitly invoking multiple linear isomorphisms, we strive to provide theorem statements, notation and simplification lemmas that let the future user ignore such complication\todo{I'm not entirely sure we do enough of this to claim it so boldly. I have some examples, but I'm not convinced you could do whatever you want without looking into the innards of the definitions.}.

The work presented here constitutes a contribution to the area of formalisation in itself.
It is also part of a larger project striving for the formalisation of theories in physics, many of which require richer mathematical context than is currently available.
Lie groups arise as symmetry groups of numerous geometric objects: as such, they find many applications in the natural sciences.
Consequently, our formalisation of Lie groups and some surrounding theorems is interesting not only as a mechanised representation of mathematics, but opens up new areas to formal study in Isabelle/HOL.
One notable example is the physics of elementary particles, where the representation theory of Lie groups is essential \cite{woit2017}.
As well as contributing to an expanding library of HOL mathematics beyond undergraduate level, this formalisation aspires to enable mechanisation in theoretical physics and other sciences.

\backmatter








\bmhead{Funding}
The present project is supported by the National Research Fund, Luxembourg (AFR Grant Number 15671644, Richard Schmoetten).

\bmhead{Author contribution}
Richard Schmoetten is the primary author of this work. Jacques Fleuriot supervised all aspects of this work, provided guidance and help, and critically reviewed all versions of this manuscript.

\section*{Declarations}

\bmhead{Competing interests}
The authors have no competing interests to declare that are relevant to the content of this article.

\bibliography{refs-bibtex-url}


\begin{thebibliography}{27}
\ifx \bisbn   \undefined \def \bisbn  #1{ISBN #1}\fi
\ifx \binits  \undefined \def \binits#1{#1}\fi
\ifx \bauthor  \undefined \def \bauthor#1{#1}\fi
\ifx \batitle  \undefined \def \batitle#1{#1}\fi
\ifx \bjtitle  \undefined \def \bjtitle#1{#1}\fi
\ifx \bvolume  \undefined \def \bvolume#1{\textbf{#1}}\fi
\ifx \byear  \undefined \def \byear#1{#1}\fi
\ifx \bissue  \undefined \def \bissue#1{#1}\fi
\ifx \bfpage  \undefined \def \bfpage#1{#1}\fi
\ifx \blpage  \undefined \def \blpage #1{#1}\fi
\ifx \burl  \undefined \def \burl#1{\textsf{#1}}\fi
\ifx \doiurl  \undefined \def \doiurl#1{\url{https://doi.org/#1}}\fi
\ifx \betal  \undefined \def \betal{\textit{et al.}}\fi
\ifx \binstitute  \undefined \def \binstitute#1{#1}\fi
\ifx \binstitutionaled  \undefined \def \binstitutionaled#1{#1}\fi
\ifx \bctitle  \undefined \def \bctitle#1{#1}\fi
\ifx \beditor  \undefined \def \beditor#1{#1}\fi
\ifx \bpublisher  \undefined \def \bpublisher#1{#1}\fi
\ifx \bbtitle  \undefined \def \bbtitle#1{#1}\fi
\ifx \bedition  \undefined \def \bedition#1{#1}\fi
\ifx \bseriesno  \undefined \def \bseriesno#1{#1}\fi
\ifx \blocation  \undefined \def \blocation#1{#1}\fi
\ifx \bsertitle  \undefined \def \bsertitle#1{#1}\fi
\ifx \bsnm \undefined \def \bsnm#1{#1}\fi
\ifx \bsuffix \undefined \def \bsuffix#1{#1}\fi
\ifx \bparticle \undefined \def \bparticle#1{#1}\fi
\ifx \barticle \undefined \def \barticle#1{#1}\fi
\bibcommenthead
\ifx \bconfdate \undefined \def \bconfdate #1{#1}\fi
\ifx \botherref \undefined \def \botherref #1{#1}\fi
\ifx \url \undefined \def \url#1{\textsf{#1}}\fi
\ifx \bchapter \undefined \def \bchapter#1{#1}\fi
\ifx \bbook \undefined \def \bbook#1{#1}\fi
\ifx \bcomment \undefined \def \bcomment#1{#1}\fi
\ifx \oauthor \undefined \def \oauthor#1{#1}\fi
\ifx \citeauthoryear \undefined \def \citeauthoryear#1{#1}\fi
\ifx \endbibitem  \undefined \def \endbibitem {}\fi
\ifx \bconflocation  \undefined \def \bconflocation#1{#1}\fi
\ifx \arxivurl  \undefined \def \arxivurl#1{\textsf{#1}}\fi
\csname PreBibitemsHook\endcsname

\bibitem[\protect\citeauthoryear{Schottenloher}{2008}]{schottenloher2008}
\begin{bbook}
\bauthor{\bsnm{Schottenloher}, \binits{M.}}:
\bbtitle{A {{Mathematical Introduction}} to {{Conformal Field Theory}}},
\bedition{2}nd edn.
\bsertitle{Lecture {{Notes}} in {{Physics}}},
vol. \bseriesno{759}.
\bpublisher{Springer},
\blocation{Berlin, Heidelberg}
(\byear{2008}).
\doiurl{10.1007/978-3-540-68628-6}
\end{bbook}
\endbibitem

\bibitem[\protect\citeauthoryear{Stephani}{2004}]{stephani2004}
\begin{bbook}
\bauthor{\bsnm{Stephani}, \binits{H.}}:
\bbtitle{Relativity: {{An Introduction}} to {{Special}} and {{General Relativity}}},
\bedition{3}rd edn.
\bpublisher{Cambridge University Press},
\blocation{Cambridge}
(\byear{2004}).
\doiurl{10.1017/CBO9780511616532}
\end{bbook}
\endbibitem

\bibitem[\protect\citeauthoryear{Woit}{2017}]{woit2017}
\begin{bbook}
\bauthor{\bsnm{Woit}, \binits{P.}}:
\bbtitle{Quantum {{Theory}}, {{Groups}} and {{Representations}}}.
\bpublisher{Springer},
\blocation{Cham}
(\byear{2017}).
\doiurl{10.1007/978-3-319-64612-1}
\end{bbook}
\endbibitem

\bibitem[\protect\citeauthoryear{Bordg et~al.}{2022}]{bordg2022}
\begin{barticle}
\bauthor{\bsnm{Bordg}, \binits{A.}},
\bauthor{\bsnm{Paulson}, \binits{L.}},
\bauthor{\bsnm{Li}, \binits{W.}}:
\batitle{Simple {{Type Theory}} is not too {{Simple}}: {{Grothendieck}}'s {{Schemes Without Dependent Types}}}.
\bjtitle{Experimental Mathematics}
\bvolume{31}(\bissue{2}),
\bfpage{364}--\blpage{382}
(\byear{2022})
\doiurl{10.1080/10586458.2022.2062073}
\end{barticle}
\endbibitem

\bibitem[\protect\citeauthoryear{Immler and Zhan}{2019}]{immler2019}
\begin{bchapter}
\bauthor{\bsnm{Immler}, \binits{F.}},
\bauthor{\bsnm{Zhan}, \binits{B.}}:
\bctitle{Smooth manifolds and types to sets for linear algebra in {{Isabelle}}/{{HOL}}}.
In: \bbtitle{Proceedings of the 8th {{ACM SIGPLAN International Conference}} on {{Certified Programs}} And {{Proofs}}}.
\bsertitle{{{CPP}} 2019},
pp. \bfpage{65}--\blpage{77}.
\bpublisher{Association for Computing Machinery},
\blocation{New York, NY, USA}
(\byear{2019}).
\doiurl{10.1145/3293880.3294093}
\end{bchapter}
\endbibitem

\bibitem[\protect\citeauthoryear{Guttmann}{2020}]{guttmann2020}
\begin{bchapter}
\bauthor{\bsnm{Guttmann}, \binits{W.}}:
\bctitle{Reasoning {{About Algebraic Structures}} with {{Implicit Carriers}} in {{Isabelle}}/{{HOL}}}.
In: \beditor{\bsnm{Peltier}, \binits{N.}},
\beditor{\bsnm{{Sofronie-Stokkermans}}, \binits{V.}} (eds.)
\bbtitle{Automated {{Reasoning}}}.
\bsertitle{Lecture {{Notes}} in {{Computer Science}}},
pp. \bfpage{236}--\blpage{253}.
\bpublisher{Springer},
\blocation{Cham}
(\byear{2020}).
\doiurl{10.1007/978-3-030-51054-1_14}
\end{bchapter}
\endbibitem

\bibitem[\protect\citeauthoryear{Paulson}{1989}]{paulson1989}
\begin{barticle}
\bauthor{\bsnm{Paulson}, \binits{L.C.}}:
\batitle{The foundation of a generic theorem prover}.
\bjtitle{Journal of Automated Reasoning}
\bvolume{5}(\bissue{3}),
\bfpage{363}--\blpage{397}
(\byear{1989})
\doiurl{10.1007/BF00248324}
\end{barticle}
\endbibitem

\bibitem[\protect\citeauthoryear{Paulson}{1990}]{paulson1990}
\begin{bchapter}
\bauthor{\bsnm{Paulson}, \binits{L.C.}}:
\bctitle{A formulation of the simple theory of types (for {{Isabelle}})}.
In: \beditor{\bsnm{{Martin-L{\"o}f}}, \binits{P.}},
\beditor{\bsnm{Mints}, \binits{G.}} (eds.)
\bbtitle{{{COLOG-88}}}.
\bsertitle{Lecture {{Notes}} in {{Computer Science}}},
pp. \bfpage{246}--\blpage{274}.
\bpublisher{Springer},
\blocation{Berlin, Heidelberg}
(\byear{1990}).
\doiurl{10.1007/3-540-52335-9_58}
\end{bchapter}
\endbibitem

\bibitem[\protect\citeauthoryear{Wenzel}{}]{isar-ref}
\begin{botherref}
\oauthor{\bsnm{Wenzel}, \binits{M.}}:
The {{Isabelle}}/{{Isar Reference Manual}}.
\url{https://isabelle.in.tum.de/doc/isar-ref.pdf}
\end{botherref}
\endbibitem

\bibitem[\protect\citeauthoryear{Kamm{\"u}ller et~al.}{1999}]{kammuller1999}
\begin{bchapter}
\bauthor{\bsnm{Kamm{\"u}ller}, \binits{F.}},
\bauthor{\bsnm{Wenzel}, \binits{M.}},
\bauthor{\bsnm{Paulson}, \binits{L.C.}}:
\bctitle{Locales: {{A Sectioning Concept}} for {{Isabelle}}}.
In: \beditor{\bsnm{Bertot}, \binits{Y.}},
\beditor{\bsnm{Dowek}, \binits{G.}},
\beditor{\bsnm{Th{\'e}ry}, \binits{L.}},
\beditor{\bsnm{Hirschowitz}, \binits{A.}},
\beditor{\bsnm{Paulin}, \binits{C.}} (eds.)
\bbtitle{Theorem {{Proving}} in {{Higher Order Logics}}}.
\bsertitle{Lecture {{Notes}} in {{Computer Science}}},
pp. \bfpage{149}--\blpage{165}.
\bpublisher{Springer},
\blocation{Berlin, Heidelberg}
(\byear{1999}).
\doiurl{10.1007/3-540-48256-3_11}
\end{bchapter}
\endbibitem

\bibitem[\protect\citeauthoryear{Ballarin}{2014}]{ballarin2014}
\begin{barticle}
\bauthor{\bsnm{Ballarin}, \binits{C.}}:
\batitle{Locales: {{A Module System}} for {{Mathematical Theories}}}.
\bjtitle{Journal of Automated Reasoning}
\bvolume{52}(\bissue{2}),
\bfpage{123}--\blpage{153}
(\byear{2014})
\doiurl{10.1007/s10817-013-9284-7}
\end{barticle}
\endbibitem

\bibitem[\protect\citeauthoryear{Ballarin}{2020}]{ballarin2020}
\begin{barticle}
\bauthor{\bsnm{Ballarin}, \binits{C.}}:
\batitle{Exploring the {{Structure}} of an {{Algebra Text}} with {{Locales}}}.
\bjtitle{Journal of Automated Reasoning}
\bvolume{64}(\bissue{6}),
\bfpage{1093}--\blpage{1121}
(\byear{2020})
\doiurl{10.1007/s10817-019-09537-9}
\end{barticle}
\endbibitem

\bibitem[\protect\citeauthoryear{Haftmann and Wenzel}{2007}]{haftmann2007}
\begin{bchapter}
\bauthor{\bsnm{Haftmann}, \binits{F.}},
\bauthor{\bsnm{Wenzel}, \binits{M.}}:
\bctitle{Constructive {{Type Classes}} in {{Isabelle}}}.
In: \beditor{\bsnm{Altenkirch}, \binits{T.}},
\beditor{\bsnm{McBride}, \binits{C.}} (eds.)
\bbtitle{Types for {{Proofs}} and {{Programs}}},
pp. \bfpage{160}--\blpage{174}.
\bpublisher{Springer},
\blocation{Berlin, Heidelberg}
(\byear{2007}).
\doiurl{10.1007/978-3-540-74464-1_11}
\end{bchapter}
\endbibitem

\bibitem[\protect\citeauthoryear{H{\"o}lzl et~al.}{2013}]{holzl2013}
\begin{bchapter}
\bauthor{\bsnm{H{\"o}lzl}, \binits{J.}},
\bauthor{\bsnm{Immler}, \binits{F.}},
\bauthor{\bsnm{Huffman}, \binits{B.}}:
\bctitle{Type {{Classes}} and {{Filters}} for {{Mathematical Analysis}} in {{Isabelle}}/{{HOL}}}.
In: \beditor{\bsnm{Blazy}, \binits{S.}},
\beditor{\bsnm{{Paulin-Mohring}}, \binits{C.}},
\beditor{\bsnm{Pichardie}, \binits{D.}} (eds.)
\bbtitle{Interactive {{Theorem Proving}}}.
\bsertitle{Lecture {{Notes}} in {{Computer Science}}},
pp. \bfpage{279}--\blpage{294}.
\bpublisher{Springer},
\blocation{Berlin, Heidelberg}
(\byear{2013}).
\doiurl{10.1007/978-3-642-39634-2_21}
\end{bchapter}
\endbibitem

\bibitem[\protect\citeauthoryear{Lee}{2012}]{lee2012}
\begin{bbook}
\bauthor{\bsnm{Lee}, \binits{J.M.}}:
\bbtitle{Introduction to {{Smooth Manifolds}}}.
\bsertitle{Graduate {{Texts}} in {{Mathematics}}},
vol. \bseriesno{218}.
\bpublisher{Springer},
\blocation{New York, NY}
(\byear{2012}).
\doiurl{10.1007/978-1-4419-9982-5}
\end{bbook}
\endbibitem

\bibitem[\protect\citeauthoryear{Spivak}{1999}]{spivak1999}
\begin{bbook}
\bauthor{\bsnm{Spivak}, \binits{M.}}:
\bbtitle{A {{Comprehensive Introduction}} to {{Differential Geometry}}}
vol. \bseriesno{1},
\bedition{3}rd edn.
\bpublisher{Publish or Perish, Inc.},
\blocation{Houston, Texas}
(\byear{1999})
\end{bbook}
\endbibitem

\bibitem[\protect\citeauthoryear{Huffman and Kun{\v c}ar}{2013}]{huffman2013}
\begin{bchapter}
\bauthor{\bsnm{Huffman}, \binits{B.}},
\bauthor{\bsnm{Kun{\v c}ar}, \binits{O.}}:
\bctitle{Lifting and {{Transfer}}: {{A Modular Design}} for {{Quotients}} in {{Isabelle}}/{{HOL}}}.
In: \beditor{\bsnm{Gonthier}, \binits{G.}},
\beditor{\bsnm{Norrish}, \binits{M.}} (eds.)
\bbtitle{Certified {{Programs}} and {{Proofs}}}.
\bsertitle{Lecture {{Notes}} in {{Computer Science}}},
pp. \bfpage{131}--\blpage{146}.
\bpublisher{Springer},
\blocation{Cham}
(\byear{2013}).
\doiurl{10.1007/978-3-319-03545-1_9}
\end{bchapter}
\endbibitem

\bibitem[\protect\citeauthoryear{Adelsberger et~al.}{2014}]{Cayley_Hamilton-AFP}
\begin{botherref}
\oauthor{\bsnm{Adelsberger}, \binits{S.}},
\oauthor{\bsnm{Hetzl}, \binits{S.}},
\oauthor{\bsnm{Pollak}, \binits{F.}}:
The {{Cayley-Hamilton Theorem}}.
Archive of Formal Proofs
(2014).
Chap. entries.
Accessed 2022-08-23
\end{botherref}
\endbibitem

\bibitem[\protect\citeauthoryear{Harrison}{2013}]{harrison2013}
\begin{barticle}
\bauthor{\bsnm{Harrison}, \binits{J.}}:
\batitle{The {{HOL Light Theory}} of {{Euclidean Space}}}.
\bjtitle{Journal of Automated Reasoning}
\bvolume{50}(\bissue{2}),
\bfpage{173}--\blpage{190}
(\byear{2013})
\doiurl{10.1007/s10817-012-9250-9}
\end{barticle}
\endbibitem

\bibitem[\protect\citeauthoryear{Kappelmann}{2023}]{kappelmann2023}
\begin{botherref}
\oauthor{\bsnm{Kappelmann}, \binits{K.}}:
Transport via {{Partial Galois Connections}} and {{Equivalences}}.
arXiv
(2023).
\url{http://arxiv.org/abs/2303.05244}
Accessed 2023-09-06
\end{botherref}
\endbibitem

\bibitem[\protect\citeauthoryear{Divas{\'o}n and Thiemann}{2022}]{divason2022}
\begin{barticle}
\bauthor{\bsnm{Divas{\'o}n}, \binits{J.}},
\bauthor{\bsnm{Thiemann}, \binits{R.}}:
\batitle{A {{Formalization}} of the {{Smith Normal Form}} in {{Higher-Order Logic}}}.
\bjtitle{Journal of Automated Reasoning}
\bvolume{66}(\bissue{4}),
\bfpage{1065}--\blpage{1095}
(\byear{2022})
\doiurl{10.1007/s10817-022-09631-5}
\end{barticle}
\endbibitem

\bibitem[\protect\citeauthoryear{{de Moura} et~al.}{2015}]{demoura2015}
\begin{bchapter}
\bauthor{\bsnm{{de Moura}}, \binits{L.}},
\bauthor{\bsnm{Kong}, \binits{S.}},
\bauthor{\bsnm{Avigad}, \binits{J.}},
\bauthor{\bsnm{{van Doorn}}, \binits{F.}},
\bauthor{\bsnm{{von Raumer}}, \binits{J.}}:
\bctitle{The {{Lean Theorem Prover}} ({{System Description}})}.
In: \beditor{\bsnm{Felty}, \binits{A.P.}},
\beditor{\bsnm{Middeldorp}, \binits{A.}} (eds.)
\bbtitle{Automated {{Deduction}} - {{CADE-25}}}.
\bsertitle{Lecture {{Notes}} in {{Computer Science}}},
pp. \bfpage{378}--\blpage{388}.
\bpublisher{Springer},
\blocation{Cham}
(\byear{2015}).
\doiurl{10.1007/978-3-319-21401-6_26}
\end{bchapter}
\endbibitem

\bibitem[\protect\citeauthoryear{Bordg and Cavalleri}{2021}]{bordg2021}
\begin{botherref}
\oauthor{\bsnm{Bordg}, \binits{A.}},
\oauthor{\bsnm{Cavalleri}, \binits{N.}}:
Elements of {{Differential Geometry}} in {{Lean}}: {{A Report}} for {{Mathematicians}}.
arXiv
(2021).
\doiurl{10.48550/arXiv.2108.00484}
\end{botherref}
\endbibitem

\bibitem[\protect\citeauthoryear{Nash}{2022}]{nash2022}
\begin{bchapter}
\bauthor{\bsnm{Nash}, \binits{O.}}:
\bctitle{Formalising lie algebras}.
In: \bbtitle{Proceedings of the 11th {{ACM SIGPLAN International Conference}} on {{Certified Programs}} And {{Proofs}}}.
\bsertitle{{{CPP}} 2022},
pp. \bfpage{239}--\blpage{250}.
\bpublisher{Association for Computing Machinery},
\blocation{New York, NY, USA}
(\byear{2022}).
\doiurl{10.1145/3497775.3503672}
\end{bchapter}
\endbibitem

\bibitem[\protect\citeauthoryear{Buzzard et~al.}{2022}]{buzzard2022}
\begin{barticle}
\bauthor{\bsnm{Buzzard}, \binits{K.}},
\bauthor{\bsnm{Hughes}, \binits{C.}},
\bauthor{\bsnm{Lau}, \binits{K.}},
\bauthor{\bsnm{Livingston}, \binits{A.}},
\bauthor{\bsnm{Mir}, \binits{R.F.}},
\bauthor{\bsnm{Morrison}, \binits{S.}}:
\batitle{Schemes in {{Lean}}}.
\bjtitle{Experimental Mathematics}
\bvolume{31}(\bissue{2}),
\bfpage{355}--\blpage{363}
(\byear{2022})
\doiurl{10.1080/10586458.2021.1983489}
{\href{https://arxiv.org/abs/2101.02602}{{arxiv:2101.02602}}}
{[math]}
\end{barticle}
\endbibitem

\bibitem[\protect\citeauthoryear{Milehins}{2022}]{milehins2022}
\begin{bchapter}
\bauthor{\bsnm{Milehins}, \binits{M.}}:
\bctitle{An extension of the framework types-to-sets for {{Isabelle}}/{{HOL}}}.
In: \bbtitle{Proceedings of the 11th {{ACM SIGPLAN International Conference}} on {{Certified Programs}} And {{Proofs}}}.
\bsertitle{{{CPP}} 2022},
pp. \bfpage{180}--\blpage{196}.
\bpublisher{Association for Computing Machinery},
\blocation{New York, NY, USA}
(\byear{2022}).
\doiurl{10.1145/3497775.3503674}
\end{bchapter}
\endbibitem

\bibitem[\protect\citeauthoryear{Kun{\v c}ar and Popescu}{2016}]{kuncar2016}
\begin{bchapter}
\bauthor{\bsnm{Kun{\v c}ar}, \binits{O.}},
\bauthor{\bsnm{Popescu}, \binits{A.}}:
\bctitle{From {{Types}} to {{Sets}} by {{Local Type Definitions}} in {{Higher-Order Logic}}}.
In: \beditor{\bsnm{Blanchette}, \binits{J.C.}},
\beditor{\bsnm{Merz}, \binits{S.}} (eds.)
\bbtitle{Interactive {{Theorem Proving}}}
vol. \bseriesno{9807},
pp. \bfpage{200}--\blpage{218}.
\bpublisher{Springer},
\blocation{Cham}
(\byear{2016}).
\doiurl{10.1007/978-3-319-43144-4_13}
\end{bchapter}
\endbibitem

\end{thebibliography}

\end{document}